# Effects of uncoordinated electrification on energy burdens for natural gas customers

Jaime Garibay-Rodriguez,[1,2,3*] Morgan R. Edwards,[1,2], Ann F. Fink[1], Zeyneb Magavi[4]

[1] La Follette School of Public Affairs, University of Wisconsin–Madison
[2] Nelson Institute Center for Sustainability and the Global Environment, University of Wisconsin–Madison
[3] Office of Sustainability, University of Wisconsin–Madison
[4] HEET (Home Energy Efficiency Team), Cambridge, Massachusetts
* Corresponding author: morgan.edwards@wisc.edu, garibayrodri@wisc.edu

**Abstract**

Electrification presents opportunities and challenges for household energy affordability. Without careful coordination, the costs of natural gas systems will be borne by a shrinking customer base, driving up bills for those left behind. This affordability issue is worsened by new pipeline investments, which risk locking communities into carbon-intensive infrastructure. Here, we introduce a framework to quantify the effects of uncoordinated electrification on natural gas customers, integrating household energy data with utility financial and planning documents. Applying our framework first to Massachusetts and then nationwide, we show that vulnerable communities face disproportionate affordability risks during heating electrification transitions. Households that do not electrify may bear a 46% increase in energy bills on average over the next 15 years (relative to current costs). Targeted electrification may help to alleviate energy burdens, but an effective heating transition will ultimately require coordinated, neighborhood-scale strategies that consider the high fixed costs of legacy infrastructure.

# Introduction

Simultaneously advancing energy affordability and greenhouse gas reduction goals is a core challenge for climate policy [1–3]. Across the U.S., over one in four households are energy insecure, with disproportionately higher rates for households of color [4–6]. These households may forgo basic necessities to pay their energy bills or keep their homes at unsafe temperatures to reduce costs [7]. For the 47% of households that heat their homes with natural gas (and 11% using other delivered fuels) [8], fuel poverty is a particular concern, especially in cold climates, where fuels make up 35% or more of energy expenditures [9]. Residential energy use also contributes 20% of total greenhouse gas emissions, with close to one third coming from fuels burned directly in approximately one billion household appliances (e.g., furnaces, dryers, and stoves) [10] Additionally, unburned natural gas is primarily methane, a potent greenhouse gas [11], and emissions throughout the supply chain (including local distribution leaks) have adverse impacts on climate [12,13] while also posing risks to safety, health, and air quality.

There are a variety of policy levers to reduce the climate impacts of residential heating. Over the past decade, many states have focused on repairing or replacing leaky distribution pipelines [14]. Although often driven by safety concerns, fixing leaks (especially super-emitters [15,16]) is also a climate solution and provides local economic benefits (via reductions in lost gas and tree damage, which impacts property values [17]). However, repairs may not be effective at permanently eliminating leaks [18]. Replacing pipelines, while effective [19], is expensive. Pipeline replacement programs can reduce explosion risks and thus have important safety benefits. However, long-term climate goals call for net zero emissions heat. Even if leaks were completely eliminated, they only represent a fraction (less than 10%) of the greenhouse gas impacts of heating with gas (with the majority coming from $CO_2$ emissions from combustion) [13]. While there are a several potential strategies to reach net zero emissions heat in buildings, including biogas or hydrogen fuels and offsetting emissions [20] electrification via heat pumps (and low-carbon electricity) is likely to be the most cost-effective and scalable solution [21].

There are signals that building electrification is accelerating in the U.S. (and globally) [22]. The Inflation Reduction Act (IRA) substantially expanded incentives for heat pump adoption, building on previous state, local, and utility programs [23]. Despite recent prospects for IRA repeal, building electrification will likely continue (albeit at a slower pace). However, there is evidence of emerging inequities in building electrification [24]. Cost remains a primary decision driver for adopting new technologies such as heat pumps, and thus those with greater upfront capital or access to credit may be more likely to install these technologies, even with incentive programs. At the same time, utilities have planned or recently completed major pipeline replacement programs, and state public utility commissions continue to approve major capital investments through gas system modernization programs [19,25,26]. This portfolio of policies risks exacerbating affordability challenges by simultaneously promoting uncoordinated electrification via voluntary technology adoption (or uptake of incentives) and new investments in natural gas infrastructure that are typically recovered through long-term ratepayer surcharges.

Current electrification trends, coupled with investments in new natural gas infrastructure, also raise the risk of a *utility death spiral*. This feedback loop occurs when customers leave a utility and fixed costs (which represent approximately half of total costs for gas utilities [27]) must be spread over a smaller group of remaining customers, driving up energy bills and causing more customers to leave. The utility death spiral was first discussed fifty years ago in the context of electric utilities, with initial concerned focused on the effects of the oil embargo [28] and more recent ones on the rise in distributed electricity generation [29,30]. While predictions of an electric utility death spiral have not materialized [31,32], and future electricity demand may increase in the future (with electrification, computing, etc. [33,34]), declining demand for natural gas is expected under climate policies. Previous work has shown how historical natural gas utility costs have changed with the customer base [27]. However, studies have yet to explore whether a natural gas utility death spiral may occur in the future, or how it may be prevented.

Beyond the risk of a death spiral, building electrification highlights the broader need to design climate policies for the mid-transition [35], a period of time (likely decades) when both fossil fuel and low-carbon energy systems will coexist at a sufficient scale to impose meaningful constraints on one other [36–38]. Key challenges in the mid-transition include: (1) continuity of energy services [35,39] (e.g., access to both gas stations and electric vehicle charging), (2) stranded assets [40–42], and (3) social and worker transitions [43,44]. Many studies focus on the challenges of phasing in new technologies [23,24,45,46]. Previous work on phaseout largely focuses on transition pathways [47–49], stranded assets and carbon lock-in [50–52], and fossil fuel jobs, with a large focus on coal communities [43,53]. Inequities in energy transitions are an important concern and have been observed in exposures to harms from energy infrastructure [54–56] and access to the benefits of low-carbon technologies such as rooftop solar [45,57], heat pumps [24,58] energy efficiency [46,59] and electric vehicles and infrastructure [60,61]. These inequities motivate frameworks for a just energy transition that call for a deliberate approach to the decline of the fossil fuel regime that centers on distributional, procedural, and recognition justice [36,62].

Here we develop a modeling framework to assess the affordability impacts of uncoordinated fossil fuel transitions and apply it to natural gas distribution systems. Our approach captures the simultaneous effects of pipeline replacement and heating electrification on costs for households that have not electrified. While pipeline replacement programs address important safety concerns, particularly for households disproportionately exposed to gas leaks, these investments can also drive up the costs of gas services, presenting a difficult tradeoff between different forms of energy injustice. We aim to address three questions with a justice lens:

1. How does the distribution of new pipeline infrastructure vary across communities?
2. How do the costs of natural gas services change throughout the heating transition?
3. What are the effects of these changing gas costs on household energy burdens?

We first apply our framework to a case study in Massachusetts and then extend it nationwide. We find that the per customer costs of natural gas in Massachusetts could increase over the next 15 years by 9% with planned pipeline replacements, 43% with building electrification, and 60% under both policies. Nationwide, for utilities with leak-prone infrastructure, combining replacement programs with electrification could increase costs by 40-57% (compared to 5% without electrification). Targeted electrification for low-income households can partially mitigate increases in energy burdens but does not address the long-term risks of a utility death spiral.

# Results

## Natural gas transitions modeling framework

Our energy transitions framework includes four major elements: (1) policies and actions, (2) cost components, (3) household characteristics, and (4) outcome metrics (see Fig. 1a). We use this framework to explore the effects of natural gas lock-in during building electrification transitions. Decisions by households to adopt electric appliances (influenced by incentive programs and other policies) will reduce demand for natural gas. This change in demand will in turn affect different components of total natural gas system costs (to varying degrees, illustrated by the grey arrows). Investments in new pipeline infrastructure will further increase system costs. How these changes impact outcome metrics such as utility bills and energy burdens will depend on the characteristics of households that do (and do not) electrify. Our framework represents these dynamics, calculating outcomes over time through 2050, with particular attention on impacts across the mid-transition and towards the end of the transition under widespread household electrification. Future work can expand this framework to incorporate links between gas and electricity infrastructure, including changing electricity costs under heating electrification and the potential for coordinated electrification strategies to reduce transition costs.

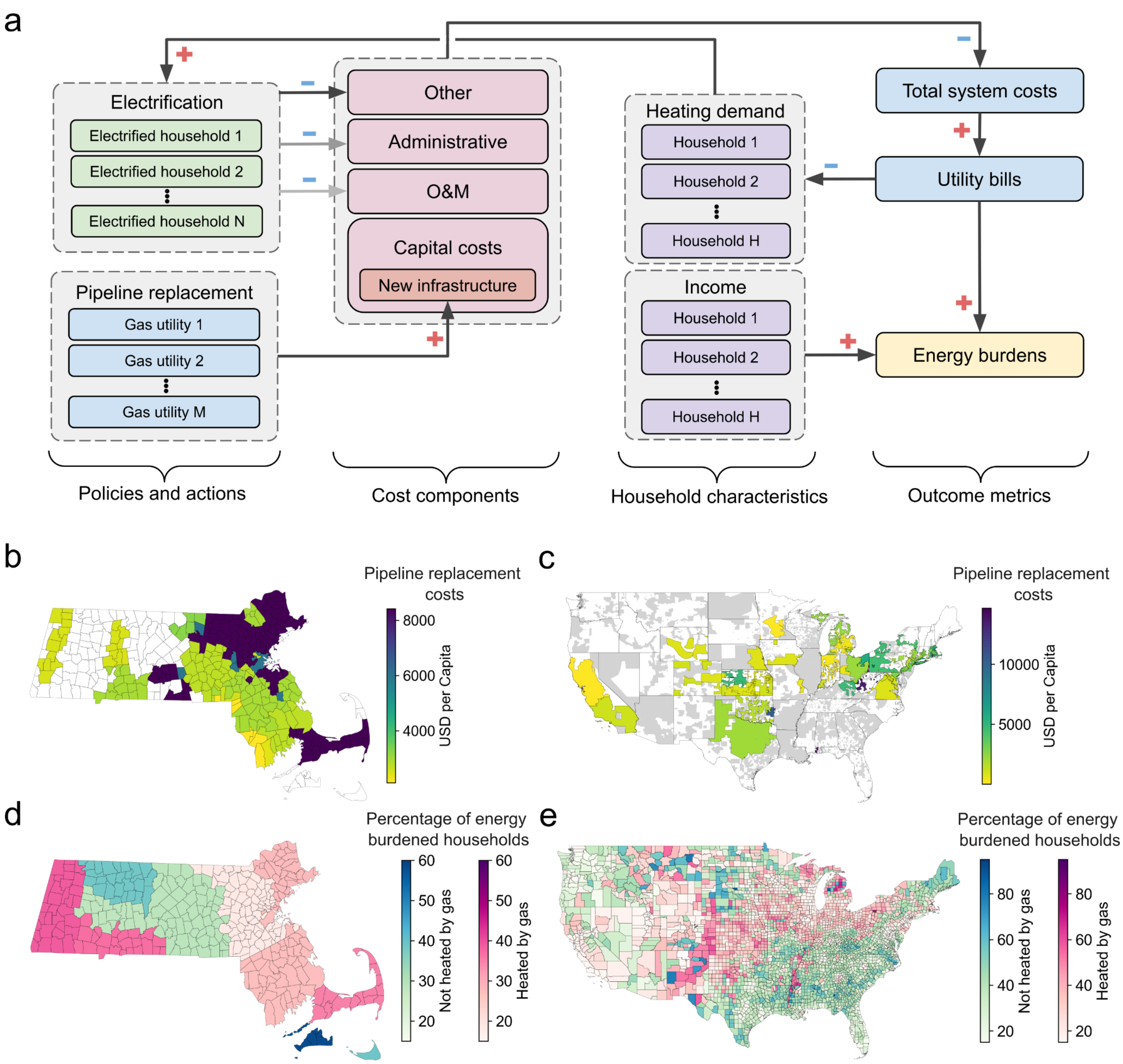

*Figure 1. Framework for modeling energy burdens from uncoordinated natural gas transitions. (a) Our modeling framework considers the combined effects of pipeline replacement and electrification policies on energy burdens for households heated with gas. Costs of replacing leak-prone pipelines (b) in Massachusetts and (c) nationally are calculated using data from the Pipeline and Hazardous Materials Safety Administration (PHMSA)* [63]. *The percentage of households that are energy burdened* [64] *(defined as spending at least 6% of household income on energy costs) and use (pink shading) or do not use (green shading) gas for heating (d) in Massachusetts and (e) nationally are calculated from the Low Income Energy Affordability Data (LEAD) Tool* [65]. *(Areas in gray shading in (c) are gas utility territories not considered in our top 50 utilities with leak-prone pipelines.)*

We apply our framework in Massachusetts (see Fig. 1b and 1d), a state with an active pipeline replacement program and ambitious climate goals. Using spatial data on natural gas infrastructure and socioeconomic indicators, we first assess inequities in pipeline replacement plans (see *Pipeline replacement patterns* in *Methods*). We then quantify potential increases in the per customer costs of gas services, considering both pipeline investments and household electrification, relative to a baseline scenario with constant costs (see *Natural gas phasedown* in *Methods*). Utilities use different depreciation methods to recover capital investments, affecting customer bills over time. We apply a straight-line depreciation approach in our main analysis and also compare it to a units-of-production method in the Supplementary Information. Using our cost projections, we assess potential impacts on energy burdens for all households currently served by gas utilities (see *Energy burden impacts* in *Methods*). Finally, we then extend our analysis nationwide (see Fig. 1c and 1e), evaluating the potential effects on per customer costs and energy burdens if the top 50 utilities (by miles of leak-prone pipelines) adopt similar policies to those in Massachusetts.

We focus our analysis on changing costs for gas customers under uncoordinated electrification and natural gas infrastructure replacement policies. However, we also note that changing electrification patterns can affect the costs of electricity services and thus home heating costs for households that electrify. We assume that heating costs for households that adopt heat pumps remain the same as current costs (i.e., costs when these households are heated with gas; see *Discussion and policy implications* for further discussion). While we do not comprehensively explore policy solutions to the gas utility death spiral, we illustrate the impacts of prioritizing early electrification of low-income households and highlight potential pathways to reduce system costs, such as neighborhood electrification in areas with aging gas infrastructure. Future work can explore these costs in more detail, and also investigate policy solutions that address interdependencies in the costs of maintaining aging natural gas infrastructure and upgrading electric grid infrastructure under increasing household electrification. See *Discussion and policy implications* for more elaboration on these solution pathways.

## Spatial patterns in pipeline replacements

Natural gas leaks are a major challenge for many utilities. As of 2022, 43,000 miles of distribution pipelines across the U.S. were classified as leak prone (i.e., cast/wrought iron and bare steel), with approximately 3,000 miles in Massachusetts [63]. These pipelines are typically old and often well beyond their intended service life. Gas leaks are disproportionately concentrated in vulnerable communities – including areas with higher representation of people of color, low-income households, and energy burdened households – both in Massachusetts and in other states with aging infrastructure [55,56]. This suggests that leak-prone pipelines are likely more common in these communities. While replacing leaky pipelines can provide safety and climate benefits, the financial burden is considerable, with costs ranging from $0.77 to $4.37 million per mile [25,26,66–68]. Massachusetts, for example, is planning to invest up to 20 billion USD to accelerate replacement of leak-prone pipelines over the next 15 years through its Gas System Enhancement Program (GSEP) [14,25]. Given these high costs, policymakers may face difficult tradeoffs between investing in immediate safety improvements through pipeline replacement and long-term solutions like electrification to advance climate goals.

We first analyze spatial patterns in pipeline replacement plans across Massachusetts [68]. Fig. 2a shows the relative percentage difference of leaks per square kilometer across population subgroups. Vulnerable populations – including (in order of disparity) households or individuals who are limited English, Black, Hispanic, low income, low education, Asian, other racial/ethnic minorities, renters, housing burdened, and disabled adults – currently experience 8-68% more leaks per area compared to the average population. Pipeline replacement density mirrors this disparity, with some vulnerable subgroups having 50% more replacements per unit area in current (2024) and near-term (2025–2028) plans (see Fig. 2b and 2c). Specifically, Hispanic and renter subgroups have over 50% more pipeline replacements in both short and long term plans, while the white and adults over 64 subgroups have consistently fewer replacements (15-25%). The cost of these investments is distributed among all ratepayers and has the positive impact of reducing gas leaks and system risks. At the same time, they present a risk of carbon lock-in and a lost opportunity to redirect investment to prioritize vulnerable communities for the benefits of decarbonization (see *Discussion and policy implications*).

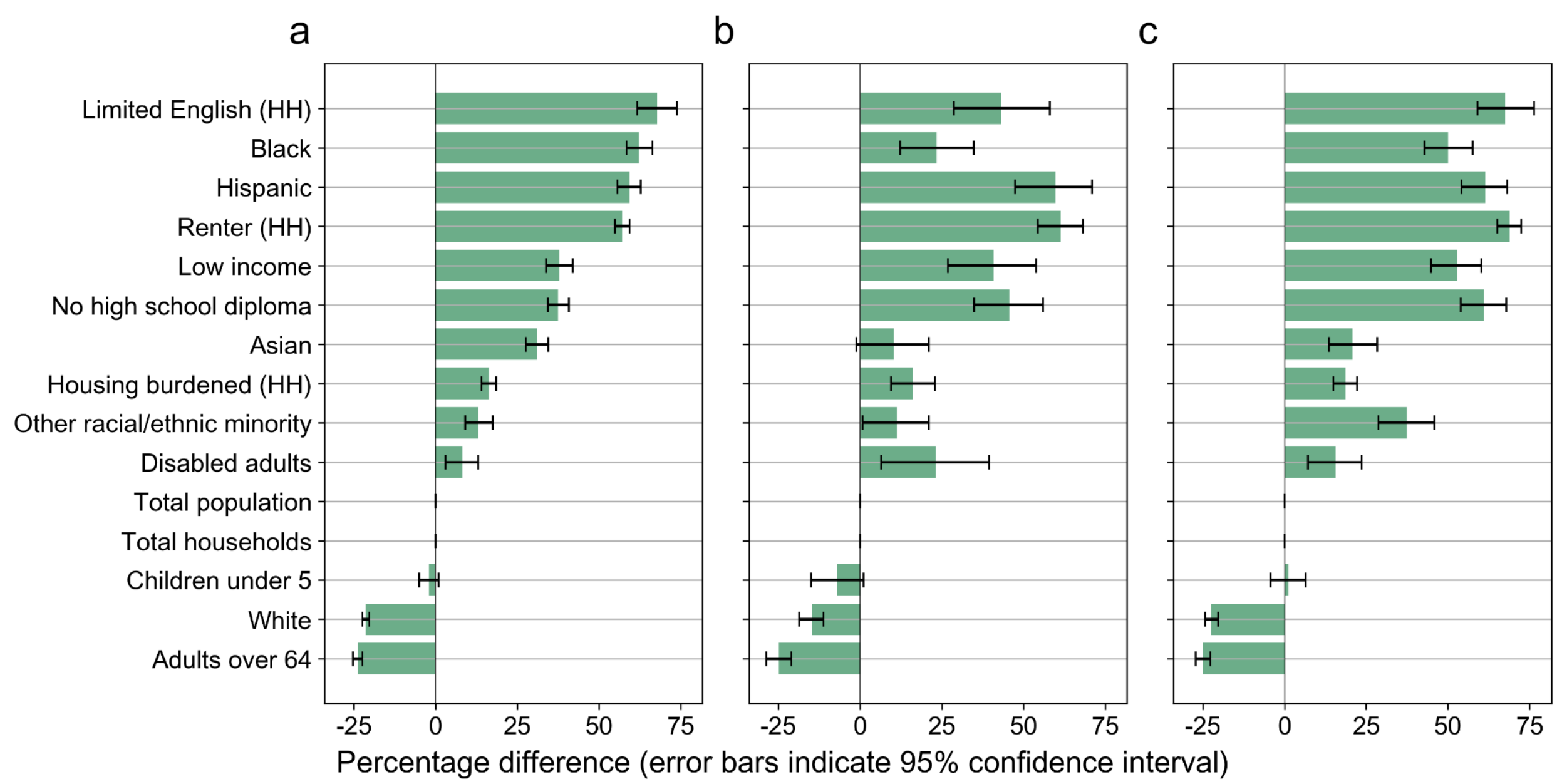

*Figure 2. Percentage difference in distribution of gas leaks and pipeline replacements for population subgroups with respect to the total population/households. Error bars show 95% confidence intervals using a Monte Carlo approach to quantify the uncertainty of the population/household estimates in each census tract (see Methods for details). (a) Natural gas leaks per unit area (square kilometer) (b) near-term pipeline replacements per unit area (plans for 2024), and (c) long-term pipeline replacements per unit area (2025–2028).*

## Rising costs of natural gas during transitions

We next model the costs of natural gas services during the energy transition. Several states, including Massachusetts, have set ambitious goals to reach net-zero emissions by 2050 [69,70]. Policies to accelerate electrification to meet these goals [23] may conflict with legacy business models for gas utilities, which rely on continued demand to recover fixed costs associated with long-lived infrastructure. We explore these dynamics under an uncoordinated electrification transition where 80% of customers electrify and thus leave the gas customer base by 2050. (Coordinated electrification may reduce these costs; see *Discussion and policy implications* for an overview of possible strategies.) Specifically, we explore three scenarios: (a) pipeline replacement, (b) building electrification, and (c) a combined policy with both replacement and electrification. We compare these to a baseline scenario where costs remain constant. Costs increase in electrification scenarios because costs per customer do not decline one-to-one when a customer leaves [27]. For example, while natural gas purchases can be fully avoided, the costs of previous capital investments must still be recovered. Other cost categories follow patterns between these two extremes (see Fig. 3a).

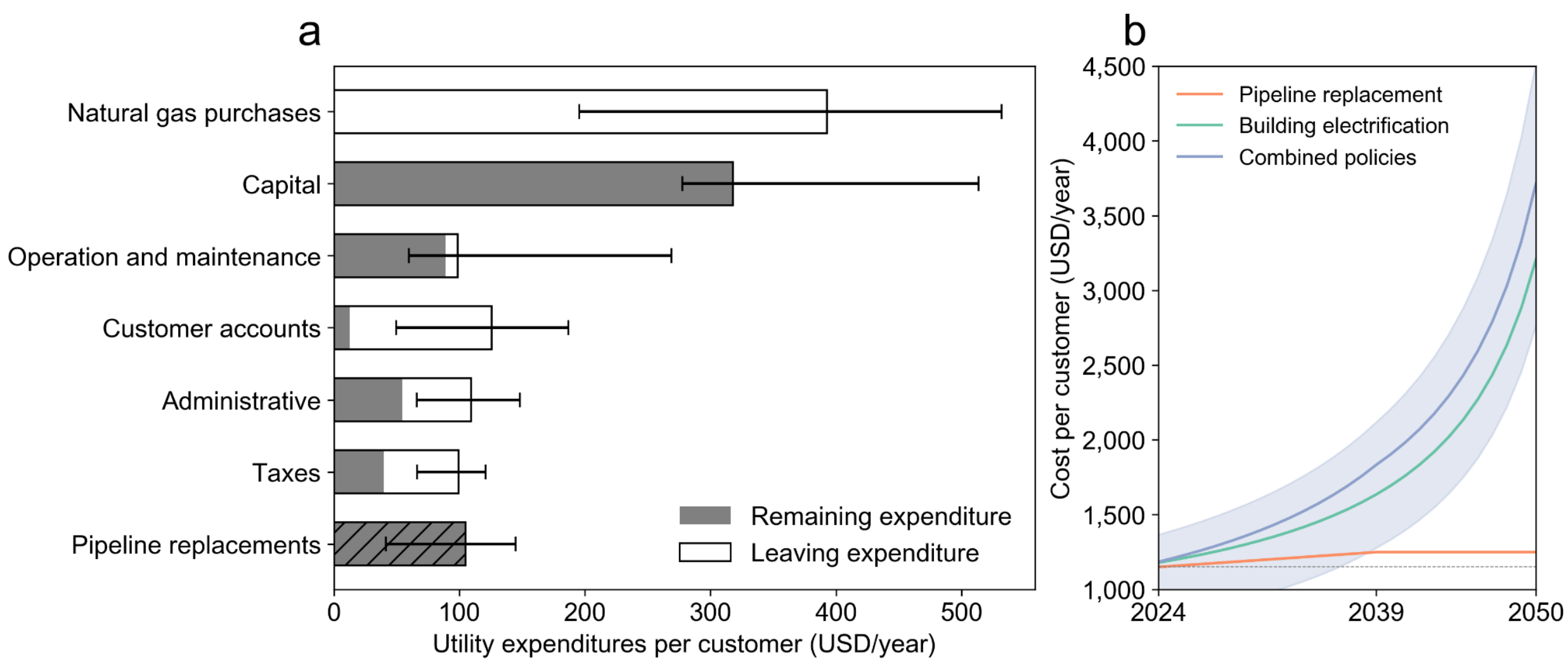

*Figure 3. Effects of an uncoordinated natural gas transition on natural gas costs in Massachusetts. (a) Customer weighted average costs per customer for each utility expenditure category. Pipeline replacement costs represent the annualized costs of all pipeline replacements by 2039. Error bars indicate the range of costs across all six major utilities in Massachusetts. (b) Average costs per customer for natural gas utilities under three scenarios: pipeline replacement, building electrification, and an uncoordinated transition that combines these two policies. Costs are compared to a baseline constant cost scenario (grey dotted line). The shaded area represents the range of costs across the six utilities in Massachusetts. Pipeline replacement scenarios use straight-line depreciation and an asset lifetime of 60 years. (We also model costs for the cases when 65% or 90% of customers electrify by 2050; see Supplementary Fig. S1).*

Building electrification and pipeline replacement both raise costs for natural gas customers, with the greatest increases occurring when these policies are pursued simultaneously (see Fig. 3b). Cost increases from pipeline replacement peak at 9.2% in 2039 (the year all replacements are scheduled to be completed) but remain high through 2050 as investment costs continue to be recovered. With electrification, a shrinking base raises per customer costs by an average of 43% by 2039 (and 180% by 2050). When these two policies are combined, costs increase further, by an average of 60% over the next 15 years (and up to 225% by 2050). Costs vary across utilities due to differences in pipeline infrastructure and expenditures (see Supplementary Fig. S2). A units-of-production depreciation method that pays off pipeline investments over their useful lifetime (which may be reduced under electrification transitions) leads to even higher costs through 2050 (see Supplementary Table S1 and Fig. S3 and S4). However, with straight-line depreciation, investments will not be completely paid off until 2099, presenting a serious challenge for cost recovery. Costs may also be higher if there is an escalation in replacement costs (see *Sensitivity analyses* in *Methods* and Supplementary Fig. S5).

## Effects of high gas costs on energy burdens

Higher per customer costs will likely raise natural gas bills, impacting future energy affordability. We model changes in the number of energy burdened households and average energy burden over time among current gas customers (see *Energy burden impacts* in *Methods*). One metric of affordability is the energy burden (i.e., the percentage of income spent on energy bills). Households are classified as energy burdened if this percentage is above a fixed threshold (e.g., 6%). (We recognize that these metrics do not fully capture all dimensions of energy burdens; see [71] for a recent review of energy justice metrics.) Using these definitions, the average energy burden (weighted by occupied housing units in LEAD, see Supplementary Note S1) among gas using households in Massachusetts is 5.7% (and 411,000 households are energy burdened). We calculate changes in energy burdens under gas system transitions using estimates of the costs of gas services and detailed data on household income and energy bills. Energy burdens depend on which customers disconnect from gas service first. We explore three illustrative scenarios: (1) customers exit in order of income, highest to lowest, (2) customers exit in reserve order of income, and (3) customers exit randomly.

If income does not affect electrification decisions (i.e., random electrification), the number of energy burdened households rises through 2045, then declines slightly by 2050 (see Fig. 4a). (This number increases again after 2050, but the gas system is unlikely to remain viable with a very small customer base; see Supplementary Fig. S6.) By 2050, 465,000 households are energy burdened, increasing to 474,000 if electrification is combined with pipeline replacement policies. Average energy burdens also rise steadily (see Fig. 4b), reaching 6.7% and 6.9% by 2050 under these two scenarios. Most households that cross the 6% threshold by 2040 are lower income (see Fig. 4c); by 2050, higher-income households begin to cross in greater numbers, while lower-income households face even greater burdens. (Some previously burdened households also no longer meet the threshold after electrifying; see Fig. 4f.) These effects are far more pronounced if higher-income households electrify first. The number of energy burdened households peaks in 2041 (513,000 and 540,000 with and without pipeline replacements, with average energy burdens of 6.7% or 7.1%), then drops sharply by 2050. However, average energy burdens in 2050 remain high due to the influence of the lowest income households, who have significantly higher energy burdens (see Fig. 4d and g).

Policies targeting low-income households for early electrification dramatically reduce (but do not eliminate) the increase in energy burdens. The number of energy burdened households peaks in 2033 (438,000 or 450,000 with and without pipeline replacements, with average burdens of 5.8% and 5.9%), then drops to near current levels by 2040 (and remains at those levels through 2050). The initial rise in energy burdens in this scenario is driven by moderate income households, with most households shifting to above the 6% threshold having a gross income of 100-150 thousand USD (see Fig. 4e). By 2050, fewer households are classified as energy burdened, but they are much higher income (above 200 thousand USD; see Fig. 4h). (Note that a relatively large mass of high income households also increase their energy burdens but do not cross the 6% threshold; see Supplementary Fig. S6). These higher-income households may also have a greater capacity to bear large increases in energy bills, illustrating the limitations of the fixed percentage approach to classifying households as energy burdened [72]. Additionally, the gas system would likely be untenable at this stage (or earlier) due to the small customer base.

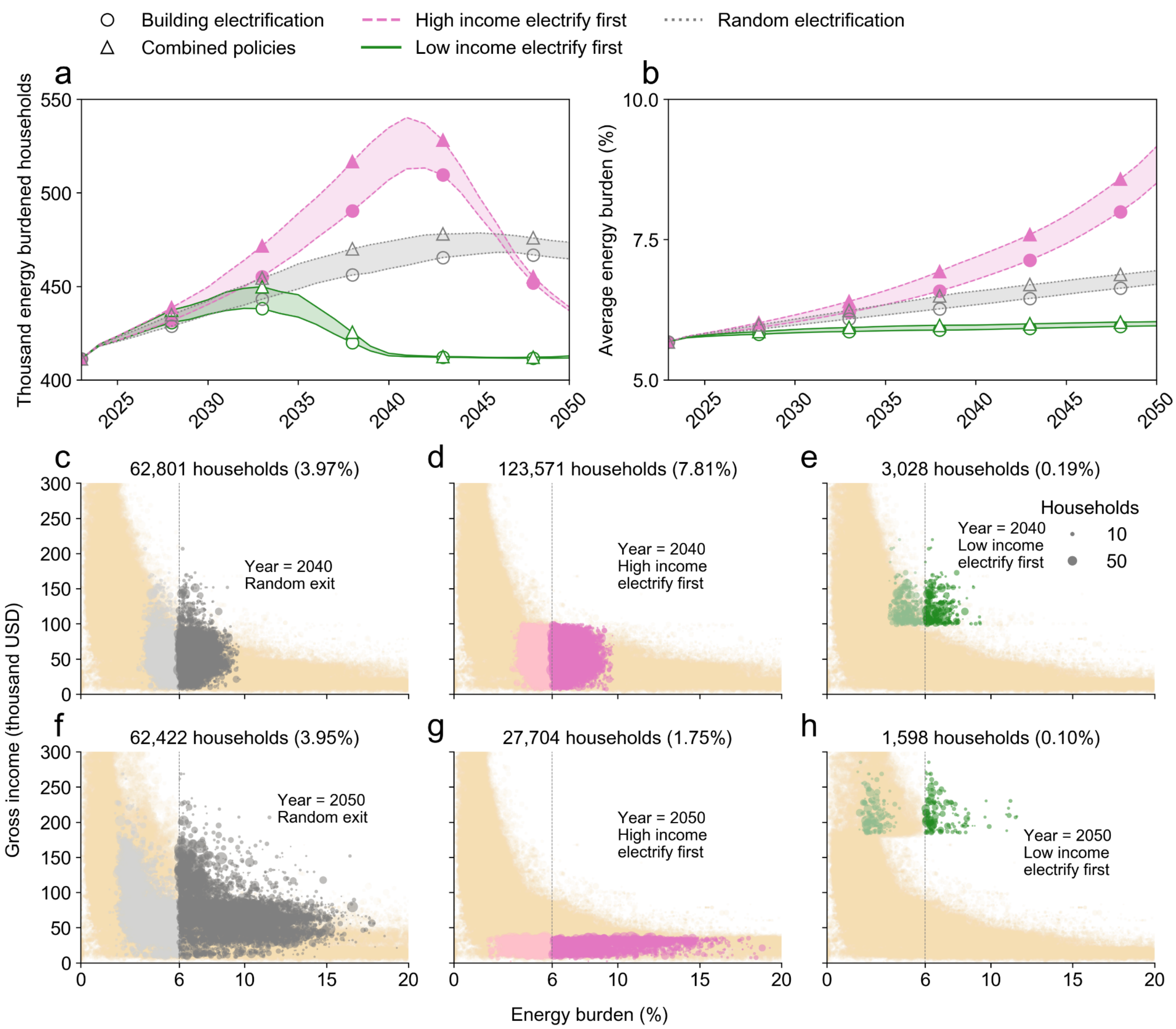

*Figure 4. Effects of an uncoordinated natural gas transition on energy burdens in Massachusetts. (a) Number of energy burdened households over time under different electrification scenarios; the shaded area represents the additional impacts of pipeline replacement policies on burdens. (b) Corresponding average energy burdens. (c)–(h) Household-level energy burdens in 2040 and 2050. Dark colored bubbles represent households*

*that become energy burdened; light colored bubbles show current energy burdens for these same households. Tan colored bubbles show households that do not cross the 6% energy burden threshold (i.e., their status as energy burdened is unchanged). Bubble size reflects the number of occupied housing units in the LEAD dataset.*

## National burdens of uncoordinated transitions

We apply our framework for modeling the effects of uncoordinated electrification transitions to the top 50 gas utilities in terms of miles of leak-prone pipelines (which represent 92% of all leak-prone pipelines in the U.S.) [63]. We focus on the potential variation in costs across utilities if they adopt policies similar to Massachusetts, using average costs per customer to estimate how costs may change as gas customers exit via electrification (see Table 2 [27]). Specifically, we model a scenario where all utilities replace their leak-prone pipelines over the next 15 years. We determine an average cost-per-mile of pipeline replacement using an internet data search (see Supplementary Table S2) and also include a sensitivity analysis using the minimum and maximum cost values (see *Sensitivity Analyses* in *Methods* and Supplementary Fig. S7). With pipeline replacement, per customer costs increase by 5% on average over the next 15 years (with a maximum of 37% across utilities) compared to current costs. With electrification, per customer costs increase by 35% over the next 15 years and 159% by 2050. With simultaneous electrification and pipeline replacement, there is substantial variation in per customer costs across utilities, with increases from 36-105% over the next 15 years (and 36-187% when considering uncertainty in replacement costs; see Fig. 5a and b).

Escalating costs of gas service can impact energy burdens across the 50 utilities. The current average energy burden is 4.7% (with a range of 2.9-9.0%) across the 37 million households serviced by these utilities (with 7.7 million households energy burdened). We present results for future energy burdens under a scenario where households electrify in random order (see Fig. 5c) and also explore bounding cases where customers exit in order of income or the reverse (see Supplementary Fig. S8). Average energy burdens increase by 0.4-1% (percentage points) by 2050 with electrification and 0.4-5% if pipeline replacement occurs simultaneously. (The effects of pipeline replacement alone, without electrification, are small.) Energy burdens are most concentrated in the Northeast and Midwest due to a larger presence of utilities with leak-prone pipelines and high heating demand (see Fig. 5b and d). This corresponds to 8.8 million energy burdened households under the combined policy (see Fig. 5c). If high income households electrify first, energy burdens increase further, peaking at 1.2% on average (with a range of 0.5-12.5%). Prioritizing low-income customers for electrification substantially mitigates but does not eliminate the increase in energy burdens, with average burdens peaking at 0.08-0.2% across utilities and an average increase of 0.1% (see Supplementary Table S6).

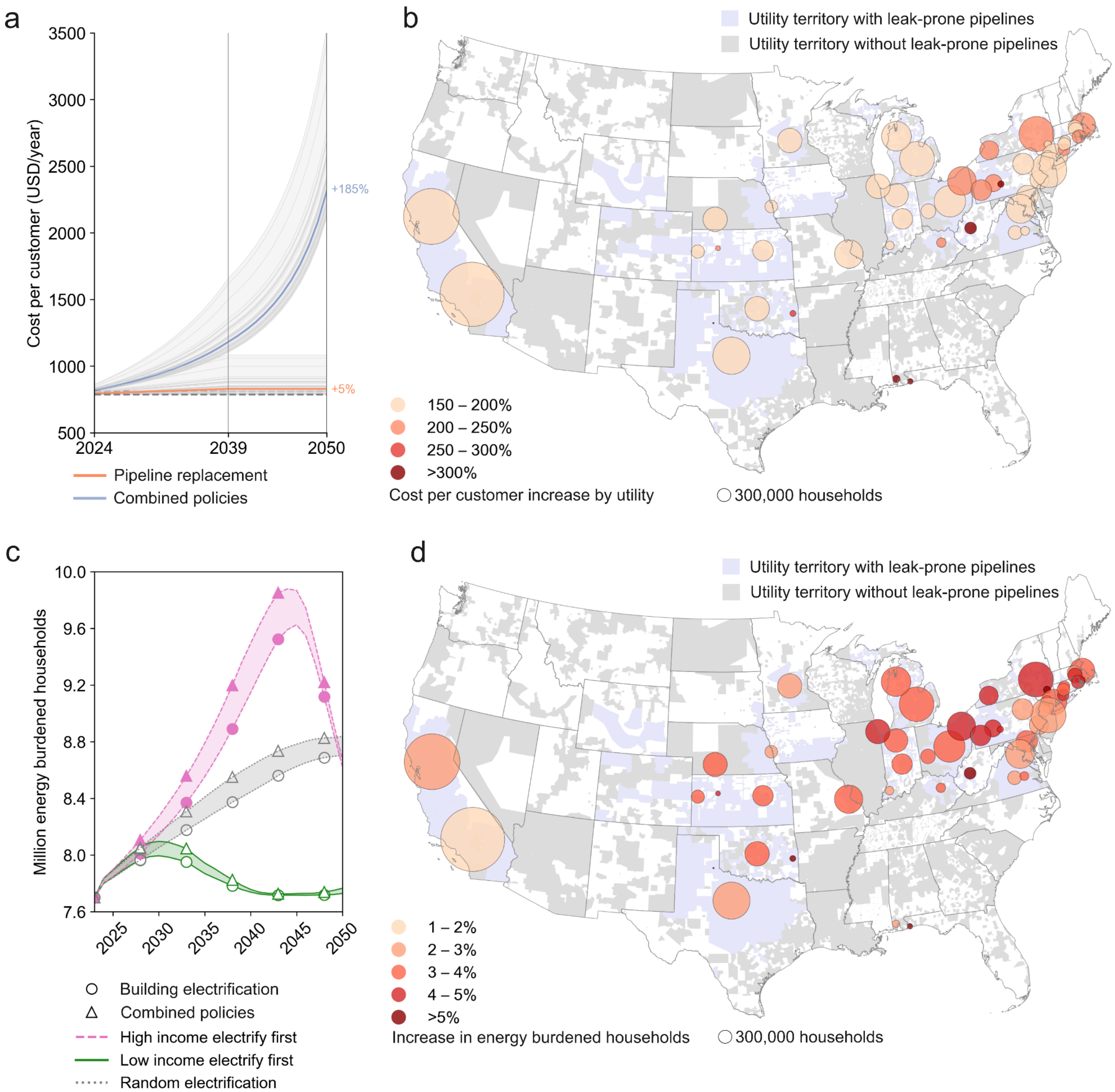

*Figure 5. National impacts of uncoordinated electrification and natural gas transitions. (a) Costs per customer for the top 50 utilities with leak-prone gas pipelines across the U.S. (b) Map of utility territories and cost per customer increases in 2050. (c) Number of energy burdened households over time under different electrification scenarios for the top 50 utilities; the shaded area represents the additional impacts of including pipeline replacement. (d) Map of the percentage increase of energy burdened households by 2050 (with customers exiting in a random order). Gray zones in maps represent all gas utilities in the U.S. and violet zones represent the territories of the 50 utilities with leak-prone gas pipelines. Bubble size is proportional to the number of residential customers.*

# Discussion and policy implications

Transitioning away from directly burning natural gas (and other delivered fuels) in buildings is crucial for achieving climate goals in the U.S. and internationally. However, ongoing investment in fossil fuel infrastructure, coupled with uncoordinated electrification at the household level, could lead to substantial increases in energy costs for residential gas consumers. These dynamics raise concerns about energy affordability and equity during the mid-transition, an extended period where some households have electrified but others continue to rely on natural gas for heat and other essential energy services. We quantify the potential effects of uncoordinated building electrification transitions on the costs of natural gas service for households that have not electrified. Our modeling framework integrates data on natural gas utility costs and pipeline replacement plans with electrification policies and household energy and sociodemographic data. Our analysis of Massachusetts highlights how leak-prone infrastructure and planned pipeline replacements are more likely to be located near vulnerable populations, including low-income, Black, Hispanic, Asian, limited English speakers, and renters. These groups may face a disproportionate risk of being locked into natural gas systems due to these investments and lower capacity to electrify their energy use [24].

An uncoordinated, customer-by-customer natural gas phasedown via electrification, whether driven by policy or market forces, may exacerbate energy burdens in Massachusetts and nationwide. However, who transitions first is also important. For a scenario where high-income households electrify first, and pipeline replacement follows current policies, the percentage of current gas heating customers in Massachusetts that are energy burdened can increase from 26% to a maximum of 34% in 2041. For the top 50 utilities nationally with the most leak-prone infrastructure, these percentages may increase from an average of 21% (and range of 9-49%) today to 24% (and range of 11-59%) across utility service areas by 2050. In contrast, a scenario that prioritizes early electrification for low-income households leads to a smaller increase (around 1%) in energy burdens, which peak in 2030 and decline through 2050. However, this household-level solution does not fully mitigate the risks of rising costs during building electrification transitions. By 2050, per customer costs for gas utilities could reach 3.4 times current levels in Massachusetts (and 2.6-4.4 times across the 50 utilities we modeled nationally). These high costs present a risk of a utility death spiral, with high potential costs for investors and society, particularly with investments in new pipeline infrastructure.

While we focus our analysis on natural gas customers, building electrification may also impact electricity costs and reliability [39]. Higher winter loads from electrified heating will require new investments in electricity systems [73]. Electrification also amplifies risks to energy service reliability, particularly during extreme weather events. While there are many strategies to mitigate these impacts, including residential energy storage [74], these solutions may increase the upfront costs of electrification and thus be distributed inequitably across households [75]. If future electricity costs decline, households that cannot electrify would be left out of potential savings, exacerbating energy injustices. However, if they increase, those same households could face compounding energy burdens, with higher per customer costs for maintaining the gas system and a share of the costs associated with electricity system expansion. Future work can address the combined effects of investments in electricity and gas systems on the costs of heating services during the mid-transition with a focus on co-optimizing solutions across these systems.

Over the next decade, policies can reduce the risk of increasing energy burdens with electrification programs for low- to moderate-income households. While its future is uncertain, the IRA offers an opportunity to support electrification for households most vulnerable to high gas costs, including point-of-sale rebates of up to $8,000 for households (or landlords) to cover heat pump installation costs. Many utilities and local governments have additional incentive programs. However, programs may vary by state, and many involve a voluntary application process and do not fully eliminate up front costs [76]. Well-designed rebate programs that minimize administrative burdens and actively seek applicants, as well as specialized training for auditors and contractors, can maximize enrollment in vulnerable communities. There are prospects for full or partial repeal of IRA, which would likely slow the pace of electrification. Moreover, the effects of a repeal would not be uniform across households, potentially exacerbating existing inequities as wealthier households with access to capital continue electrifying. Regardless of the near-term uncertainty in federal policies, the evolution of building energy technologies is underway, and the long-term risks and impacts of an unmanaged transition remain.

Continued investment in new natural gas pipelines, and other long-lived fossil fuel assets, presents a challenge for managing affordability during energy transitions. If gas utilities fail to consider future gas use (and the potential for customer exit with electrification) when planning the depreciation of new infrastructure, they risk having artificially long depreciation schedules. However, considering future transition scenarios and policies in depreciating investments in new fossil fuel assets (i.e., with the units-of-production method) may lead to higher customer costs and energy burdens in the near term. These burdens are compounded by continued pipeline replacements. These tradeoffs highlight the importance of understanding evolving energy burden dynamics (i.e., how gas system legacy costs are shifted to households with different characteristics) during electrification transitions to ensure that all households have access to the benefits of climate policy and no households are left behind. In the context of pipeline replacement, utilities can focus on strategically repairing high-volume leaks, which is more cost effective, rather than fully replacing pipelines. Methods to address leakage and safety that do not create affordability challenges should be identified and prioritized. However, repair failures can be common, indicating the importance of careful monitoring of repair outcomes [18].

There are several technology and policy solutions that may help accelerate beneficial electrification and alleviate the risks of the gas utility death spiral. At the household scale, reducing up front and operating cost of air source heat pumps (potential with addition of thermal energy storage systems [74]) may provide technological pathways that align well with climate goals and promote energy affordability. Utilities, utility commissions, state energy offices, and community-based organizations can also collaborate to create "one-stop shops" for customers seeking to apply for and receive available incentives and clean heating technologies. Whole home electrification upgrades or on-bill financing can address other cost barriers, particularly in colder climates and older, less efficient buildings. Focusing on electrifying all end uses can also help save customers on fixed charges associated with the use of the gas system. Neighborhood-scale solutions (e.g., neighborhood electrification via air source heat pumps or networked geothermal systems) can also enable strategic pruning of the natural gas network to reduce fixed infrastructure costs. Large-scale community and neighborhood electrification strategies may also help address the tension between safety and equity, mitigating gas leaks and hazards while preventing carbon lock-in and resulting affordability risks.

There are prospects for increased coordination of building energy transitions across the U.S. State policymakers, utility commissions, and utility companies will each serve important and unique roles in this process. Beyond our case study state of Massachusetts, other states are working to align gas utility regulations and policy with climate goals. For example, Nevada and Minnesota require gas utilities to submit integrated resource plans (IRPs) that ensure safe, reliable service in line with greenhouse gas emissions targets, while New York and Colorado utilities are addressing anticipated gas pipeline expansions by exploring demand reduction strategies [77]. Exploring non-gas pipeline alternatives (NPAs), which encompass investments to reduce or avoid the need to build or upgrade gas delivery infrastructure, is required by utilities in several states, such as Rhode Island, California, New York, and Colorado [78]. State regulators might also consider requiring utilities to reform gas line extension allowances, which currently incentivize new gas connections at the expense of all ratepayers [79]. In addition, as of 2024, 15 states have enacted or are developing policies that encourage fuel-switching as a means to clean heating through customer incentives, either via utilities themselves, or state-funded programs like the Regional Greenhouse Gas Initiative (RGGI) [80].

Providing affordable heat for residential buildings is one piece of a much bigger policy puzzle to support energy justice. A common theme, emphasized in our results on energy burdens during natural gas transitions, is the importance of careful planning and coordination to support affordability for all households, including those that continue to rely on fossil fuels during the mid-transition. We quantify a risk that is a growing concern among policymakers and utility customers: continued investment in the natural gas system may impose high utility costs on those already spending a large portion of their income on energy. While pipeline replacement programs can mitigate methane leaks and safety risks in an aging distribution system [19], this benefit must be weighed against the potential for increased costs to consumers. Our framework can serve as a foundation for analyzing similar issues in other sectors undergoing electrification (e.g., personal transportation), especially where there may be tradeoffs between near- and long-term climate solutions. Future research might also address the extent and impact of the wealth transfer from gas to electric utilities as demand shifts toward electricity. Utilities and regulatory commissions in particular can be key actors in managing the energy system with an eye to addressing both traditional metrics (e.g., safety, affordability, and reliability) and broader societal issues, including health, employment, and quality of life.

# Methods

Our modeling framework for assessing the equity impacts of natural gas transitions and uncoordinated electrification consists of three steps: (1) investigate spatial patterns in natural gas pipeline replacement programs, (2) model the effects of electrification and pipeline replacement policies on the costs of natural gas service, and (3) estimate the effects of these cost increases on energy burdens across households. For each of these steps, we first describe our data and methods for our application case in Massachusetts and then discuss how we extend our approach nationwide to utilities with leak-prone infrastructure

Massachusetts presents a compelling case study for modeling energy burdens during household heating transitions and the effects of carbon lock-in from new natural gas infrastructure. It is a national leader in climate policy, with a target to reach net zero emissions economy-wide by 2050, and also has plans to fully replace its network of approximately 3,000 miles of aging pipeline [63] by 2039 through the Gas System Enhancement Program (GSEP). We first analyze spatial and sociodemographic patterns in gas leaks and planned pipeline replacements in Massachusetts, using annual utility leak reports [81] and plans filed under GSEP [68] as well as data from the American Community Survey (ACS; see *Pipeline replacement patterns*) [82]. Next, we model potential increases in per customer costs under a current policies scenario (including pipeline replacement and climate goals) and assess the impacts on household energy burdens (see *Natural gas phasedown* and *Energy burden impacts*). We use data from the U.S. Pipeline and Hazardous Materials Safety Administration (PHMSA), annual utility financial reports, and energy and demographic data from the Low Income Energy Affordability Data (LEAD) Tool (see Fig. 1b and 1d) [65].

Uncoordinated electrification can also have impacts on energy bills nationwide. The IRA provides incentives for households to adopt efficient electric appliances (e.g., heat pumps) that replace natural gas and other delivered fuels [83]. Notably, 25 governors have committed to installing 20 million heat pumps by the end of the decade [70], while 24 states have set net-zero goals, collectively representing 40% of U.S. gross domestic product (GDP) [84]. We extend our modeling framework to consider the top 50 gas utilities nationwide in terms of miles of leak-prone pipelines (e.g., cast/wrought iron and bare steel; see *Natural gas phasedown*). These utilities, which collectively account for 92% of all leak-prone distribution pipelines, could initiate or have already begun replacement programs [63]. Integrating data from PHMSA and LEAD with national utility cost estimates (see Fig. 1c and 1e) [27], we explore a hypothetical scenario where utilities replace pipelines at the same rate targeted in Massachusetts (i.e., complete replacement by 2039). Among the 28 states served by these utilities, 15 have net-zero or other climate targets [69]. Following the approach in the Massachusetts case, we quantify the potential impacts on utility costs and energy burdens.

## Pipeline replacement patterns

We first assess spatial patterns in natural gas leaks and pipeline replacement plans in Massachusetts using data compiled by HEET [55,68]. Replacement plans are reported by utilities annually for the upcoming year under the Gas System Enhancement Plan (GSEP) with details on the location, estimated cost, and prioritization of each pipeline segment. HEET extracts and geocodes replacement plans for each utility [68]. The approach uses automated report data extraction with data parsers and geocoding of each GSEP location. However, reports are not required to have a specific format. Due to data format inconsistencies, HEET performs an extensive manual verification of results to maximize accuracy. Results are published as an interactive map [68] with underlying data available upon request. The data we use in our analysis reflects data published by HEET as of August 15, 2024, which includes detailed near-term pipeline replacement plans for 2024 and gas leaks and repairs through the end of 2023. We also use data on potential locations for longer-term pipeline replacement plans (i.e., 2025-2028) communicated by utilities in their annual GSEP reports.

We aggregate leak and pipeline replacement data to the census tract scale to assess exposures across different populations. Sociodemographic variables (at the census tract level) are extracted from the American Community Survey (ACS) five-year estimates for 2018-2022 [82]. We use data on race, ethnicity, low income status (i.e., below two times the poverty line), adults over 64, children under five, adults with disabilities, adults without a high school diploma, renter status, English proficiency, and housing burdens (i.e., 30% or more of income spent on rent or mortgage). We focus on three metrics: (1) gas leak density (i.e., leaks per unit area, which has also been explored in prior work [54,55]), (2) near term replacement density (i.e., number of replacements per unit area), and (3) long term replacement density. We calculate population weighted means for each metric across different subgroups:

$$\underline{\mathrm{Metric}}(\mathrm{subgroup}) = \frac{\sum_{c=1}^{C} \mathrm{Metric}_c \mathrm{Population}_c(\mathrm{subgroup})}{\sum_{c=1}^{C} \mathrm{Population}_c(\mathrm{subgroup})}, \quad (1)$$

where *C* is the number of census tracts. We then calculate the percentage difference for each metric for all population subgroups to the total population:

$$\mathrm{MetricPercentageDifference}(\mathrm{subgroup}) = 100\left(\frac{\underline{\mathrm{Metric}}(\mathrm{subgroup})}{\underline{\mathrm{Metric}}(\mathrm{total\ population})} - 1\right). \quad (2)$$

Equation (2) returns a positive value if the phenomenon is overrepresented among the subgroup relative to the population as a whole (and vice versa). It is commonly used in studies mapping inequities in exposure to environmental burdens [55,85,86].

Data in ACS is estimated based on a rolling sample of responses (~3.5 million each year) and is subject to uncertainty. We use a Monte Carlo analysis to quantify this uncertainty by constructing random normal distributions with the ACS estimates and published margin of error (MOE) for each variable. We simulate 10,000 realizations with random samples of each population estimate distribution per census tract. Note that smaller units of analysis, such as census blocks or block groups, have larger MOEs. We perform our analysis at the census tract level as a compromise that balances geographic resolution and accuracy.

## Natural gas phasedown

We model a simplified scenario for natural gas phasedown in buildings. Decarbonization goals can be met in multiple ways but will almost certainly involve electrification of energy end uses. In Massachusetts, utilities are required to submit plans to ensure compliance with state climate targets (including at least 85% decarbonization by 2050) through Order 20-80 starting in 2025 [87]. However, the exact dynamics of how this might unfold are uncertain. Plans could adopt a range of strategies at different scales, including hybrid and full electrification via air source heat pumps, networked geothermal, and decarbonized gas use with increased equipment efficiency (e.g., biogas and hydrogen) [88]. We focus on a baseline scenario where 80% of residential customers are disconnected from the network by 2050, following previous research [27], and explore other scenarios in our sensitivity analysis. We simulate this scenario by assuming that the rate at which customers leave the system is constant (~3% of current customers annually). This simplified modeling choice is not meant to fully capture the complex, nonlinear dynamics of electrification, influenced by many technical, social, and economic factors [24,58,89].

We model the effects of customer exit via electrification on the costs of natural gas service, including gas purchases, depreciation, return on net utility plant, operation and maintenance (O&M), account and administrative expenses, and taxes. For Massachusetts, we use annual reports on costs, revenues, and customer numbers from the six major utilities [81]. Costs for each category are allocated to residential service based on the proportion of sales revenue from residential customers (60% on average across utilities). We convert revenue to expenditures using a cost recovery factor (i.e., funds a utility is allowed to collect after covering expenses):

$$\text{Expenditures} = \text{Revenues} \cdot \text{CostRecoveryFactor}. \tag{3}$$

Our approach implicitly assumes that the fraction of expenditures for residential customers remains constant over time, which might be reasonable if demand from residential and non-residential customers declines at similar rates. We calculate expenditures per customer by category by dividing each category by the number of customers. We do not model rate structures by customer type; instead, we estimate the average cost per customer. See Table 1 for our estimates for residential expenditures per customer by category.

We treat pipeline replacements as a separate utility expenditure category to highlight how ongoing pipeline replacements could affect utility finances and energy bills. Using PHMSA data, we estimate the miles of leak-prone pipelines (i.e., cast/wrought iron and bare steel) for each utility [63]. We model a scenario where all remaining leak-prone pipelines are replaced by 2039, the scheduled end of the Massachusetts replacement program, with a constant replacement rate each year (see Supplementary Note S2). Replacement costs per mile (and replacement plans, shown in Fig. 2) are sourced from HEET [68] (see Supplementary Table S3 for details). For our baseline scenario, we assume no cost escalation. We also consider a scenario with a 2% annual escalation rate (see Sensitivity Analysis) [88]. Total yearly costs are:

$$\text{TotalReplacementCost} = \text{Miles} \cdot \text{CostPerMile}. \tag{4}$$

Note that because replacement costs are annualized (see description below), total annual payments for pipeline replacement increase over time until all pipelines are replaced, and payments continue even after the replacement period ends.

Utilities often choose different depreciation methods in their rate-making process to annualize the costs of capital investments (i.e., spread the cost of an asset over multiple accounting periods) [90]. There are three key regulatory principles for asset depreciation: (1) economic efficiency, (2) stability and intergenerational equity, and (3) administrative simplicity. The most common depreciation method, and the one used in previous studies, is straight-line depreciation (SLD in Equation 5), which estimates costs as a function of time [88,90,91]:

$$\mathrm{SLD_t} = \frac{\mathrm{TotalAssetCost-\ ResidualValue}}{\mathrm{T}}, \quad \mathrm{t} = 1,2,\ldots,\mathrm{T}, \tag{5}$$

where $T$ represents the asset lifetime. Another approach, units-of-production depreciation (UOP in Equation 6), depreciates an asset based on its usage over its lifetime:

$$\mathrm{UOP_t} = \left(\frac{\mathrm{TotalAssetCost-\ ResidualValue}}{\mathrm{U}}\right)\mathrm{u_t}, \quad \mathrm{t} = 1,2,\ldots,\mathrm{T}', \tag{6}$$

where $U$ is the expected number of units produced over the asset's useful life $T'$ (which may differ from the straight-line depreciation lifetime), and $u_t$ is the units produced each year.

We calculate depreciation using both the straight-line and units-of-production methods. For the straight-line depreciation method, we use an asset life of $T = 60$ years [88]. Natural gas pipelines are typically designed for a service life exceeding 50 years (and can last up to 80 years or more) [92,93]. Note that, with this method, there is a substantial cost burden from pipeline replacements even after the gas system has been largely phased down. These costs may be difficult to recover. The units-of-production method can address this concern by allowing for accelerated depreciation of assets that are expected to be retired before the end of their useful life. (A straight-line depreciation method, in contrast, implicitly assumes that demand will remain stable.) For the units-of-production method, we use the modeled decline in gas demand to calculate units produced through 2050 and assume this trend continues until complete phaseout in 2057. We assume that the residual value of pipelines and retirement costs are zero. (We note that the depreciation expenditure includes current utility infrastructure retirement costs.) With assets subject to accelerated retirement, such as natural gas pipelines, the straight-line method leads to lower annual payments than units-of-production method through 2050.

We estimate expenditures for gas utilities in Massachusetts using (1) annual reports submitted by individual utilities and published on the Department of Public Utilities (DPU) website [94], which are broken up into several cost categories, and (2) estimates of the percentage change in cost with customer departure by cost category (see Table 1) [27]. For instance, we assume all gas purchase costs are eliminated when a customer leaves, but capital-related costs remain unchanged. We assume that only 10% of operation and maintenance (O&M) costs are eliminated, as most network maintenance continues, except for customer-specific activities (e.g., meter repair). Customer account expenses are mostly eliminated (90%), though some costs remain to account for fixed expenses and lost economies of scale (e.g., in physical meter readings). Half of administrative and general expenses are eliminated, with the rest attributed to fixed costs (e.g., pensions). We estimate 60% of tax costs are eliminated based on a weighted average of other categories. These percentages are estimates and may vary across utilities and over time. Higher fixed costs (and lower customer-dependent costs) would increase the costs of gas service for remaining customers as customers leave (and vice versa).

***Table 1.*** *Expenditure categories for natural gas utilities in Massachusetts (labeled 1-6) and fraction of costs eliminated when a customer exits. Expenditures are for 2024 and estimated from individual utility reports extracted from the Massachusetts DPU website* [94]. *The fraction of costs leaving with customers are taken from previous research* [27].

| Category | 2024 USD per customer | | | | | | Fraction leaving with customer |
|---|---|---|---|---|---|---|---|
| | 1 | 2 | 3 | 4 | 5 | 6 | |
| Natural gas purchases | 351 | 532 | 437 | 195 | 278 | 292 | 1 |
| Depreciation | 150 | 154 | 100 | 135 | 119 | 243 | 0 |
| Return on net utility plant | 183 | 130 | 205 | 165 | 159 | 271 | 0 |
| Total operation and maintenance | 106 | 95 | 60 | 92 | 269 | 79 | 0.1 |
| Customer account expenses | 164 | 99 | 50 | 51 | 72 | 187 | 0.9 |
| Administrative and general expenses | 121 | 105 | 84 | 66 | 99 | 148 | 0.5 |
| Taxes | 107 | 66 | 115 | 75 | 121 | 68 | 0.6 |
| Pipeline replacements | 4 | 5 | 8 | 6 | 2 | 46 | 0 |

For our national analysis, we model a hypothetical pipeline replacement scenario where utilities with leak-prone infrastructure follow the same replacement plans as in Massachusetts. To identify leak-prone infrastructure, we use 2022 PHMSA inventories of cast/wrought iron and bare steel pipelines [63] which is the most recent year with U.S. EIA form 176 data on residential gas customers (as of November 1, 2024) [95]. We focus on the top 50 utilities with the largest number of miles of leak-prone miles of pipelines (out of 174 utilities that report inventories to PHMSA). We manually match customers reported in EIA form 176 with utility names and serviced states and merge this data with spatial data on territory boundaries from the Homeland Infrastructure Foundation-Level Data (HIFLD) [96]. After merging utility territory boundaries, we exclude five utilities form our sample due to incorrect boundaries (after visual inspection from utility websites). These excluded utilities represent 4.8% of leak-prone pipelines. To compensate, we added the next five utilities in terms of leak-prone miles to our sample, resulting in a final sample representing 50 utilities and 92% of leak-prone pipelines.

We search for cost and mileage data to determine a cost-per-mile (in 2024 U.S. dollars) for utilities within our national sample with active pipeline replacement programs. Specifically, we use utility names and search terms such as “pipeline replacement program,” “cost,” and “mileage,” reviewing the first two pages of Google results. We only consider values reported by utility websites, state utility commissions, and news outlets that cite utility representatives or reports. Each utility is assigned one of six statuses (see Supplementary Table S2):

1. Government-funded replacement program
2. No evidence of active program
3. Potential inaccuracy in mileage or cost data
4. Found one-off replacement projects but no evidence of a replacement program
5. Confirmed active replacement program but missing either mileage or cost data
6. Confirmed active replacement program

Overall, we found evidence of active or previous replacement programs for 18 of the 50 utilities and cost and mileage data for 8 of these utilities. Given limited data, we focus on a scenario using an average cost-per-mile across our utilities (1.98 million USD). We also include sensitivity analyses for replacement costs using minimum, maximum, and Massachusetts cost-per-mile values. Average utility expenditures are based on data from the American Gas Association (AGA) (as referenced in previous research [27]; see Table 2).

***Table 2.*** *Expenditure categories for a typical natural gas utility in the U.S. Data are sourced from previous research and adjusted to 2024 USD* [27]. *The fraction of expenditures leaving with customers is the same as in our Massachusetts case (see Table 1). Pipeline replacement costs per customer vary per utility due to differences in the number of customers and miles of leak-prone pipelines (see Supplementary Table S4).*

| Category | 2024 USD per customer |
| --- | --- |
| Natural gas purchases | 349 |
| Depreciation | 71 |
| Return on net utility plant | 118 |
| Total operation and maintenance | 74 |
| Customer account expenses | 28 |
| Admin and general expenses | 95 |
| Taxes | 53 |

## Energy burden impacts

We analyze the effects of changing costs of natural gas services on energy burdens using the Low Income Energy Affordability (LEAD) tool [65]. LEAD contains statistically representative household data at the census tract level including: (1) estimated number of households per census tract, (2) area median income, (3) household energy expenditures (natural gas, electricity, and other fuels), and (4) main heating fuel. In the contiguous U.S., LEAD includes approximately 16.7 million weighted data points representing 113 million households in 2018 (the most recent year for which LEAD data was available as of the end of 2024). We filter the data to consider households using natural gas for heating, which are identified in LEAD, but scale it to consider all households served by utilities. In Massachusetts, this results in average energy burdens of 5.67% and 5.73%, when including households heated by gas and all households with gas access, respectively. Using census tract identifiers, we perform a spatial merge of our utility data with the LEAD data. We obtain the counties served by each utility from the Massachusetts DPU [97] (for the Massachusetts case) and the HIFLD [96] (for the nationwide case) and information census tracts from the U.S. Census Bureau [98].

There are discrepancies between utility-reported customer counts and the total number of households using gas reported by LEAD in the utility service territory. Several factors may contribute to these discrepancies. First, the households in LEAD are based on estimates from the ACS 2018 5-year survey, whereas utility residential customer counts are based on more recent data from 2024. Second, multifamily buildings may be counted as a single gas customer by utilities if they share a gas meter but would appear as separate households in LEAD. Third, filtering households in LEAD for gas heating excludes non-heating gas customers. (Nationally, approximately 68% of gas customers use gas for heating, with this figure rising to 76% in colder climates [99].) Finally, some census tracts are serviced by more than one utility, leading to ambiguity in the number of LEAD households connected to each utility. We follow an existing approach and assign a customer count to each census tract. We scale the total number of utility customers by the fraction of LEAD households in that tract [100]:

$$\mathrm{Customers_c} = \mathrm{TotalUtilityCustomers} \cdot \frac{\mathrm{Households_c}}{\mathrm{HouseholdsInServiceArea}}, \quad \mathrm{c} = 1,2,\dots,\mathrm{C}, \tag{7}$$

where $Customers_c$ is the scaled utility customers in census tract *c*, $\mathrm{Households_c}$ is the number of LEAD households in census tract *c* that use gas for heating, and $\mathrm{HouseholdsInServiceArea}$ is the sum of these households across the utility service territory. For cases where a utility territory crosses a census tract, we multiply by the fraction of area covered by the utility.

We next calculate energy burdens for each LEAD household:

$$\mathrm{Energy\ Burden} = \frac{\mathrm{ElectricityExpenditures} + \mathrm{GasExpenditures} + \mathrm{OtherFuelExpenditures}}{\mathrm{HouseholdIncome}}. \tag{8}$$

Our estimates of increased natural gas costs, expressed as percentages relative to the initial year of our analysis (i.e., 2024), are then applied to gas expenditures for electrification scenarios with and without additional costs due to new investments in pipeline replacements, taking into account our different pipeline replacement scenarios (see Section *Effects of high gas costs on energy burdens*). Note that increases in natural gas expenditures are applied only to remaining gas customers. Electricity and other fuel expenditures as well as income remain unchanged for remaining gas customers. We conservatively assume that customers who switch from gas to electricity experience no change in total energy (i.e., electricity plus gas) costs or energy burdens. While research suggests that the majority of customers who electrify via heat pumps would see a decline in energy costs, realized costs depend on a variety of factors such as climate, electricity and gas costs, and equipment costs [24]. We also assume that real (i.e., inflation adjusted) gross (i.e., before tax) household income remains constant over time.

A crucial aspect of how electrification and pipeline replacement impact energy burdens is the composition of customers electrifying (and thus disconnecting from gas service). Current trends in heating electrification suggest a nonlinear relationship with household income moderated by variables such as local climate, energy prices, building age, and race/ethnicity [24,58,89,101]. However, other clean energy technologies like rooftop solar demonstrate adoption trends where income plays a significant role [45]. Future patterns in household electrification remain uncertain. If wealthier households electrify first, they may impose disproportionate burdens on low-income households during a gas phasedown. To assess the impact of the composition of departing customers on energy burdens, we explore two extreme scenarios – households depart in order of high to low income and the reverse – and a random electrification scenario. Our aim is to estimate the upper and lower bounds of energy burden impacts during an uncoordinated electrification. For both scenarios, we measure (1) changes in average and median energy burdens and (2) changes in the number of households that become energy burdened, using a threshold of 6% to identify energy-burdened households.

## Sensitivity analyses

We perform several sensitivity analyses for both our Massachusetts and national cases (see Supplementary Fig. S1-S8 and Tables S5 and S6 for more details):

- We model a case where 65% of customers electrify by 2050 in Massachusetts instead of 80%, which results in lower overall increases in the cost of gas service from 2024–2050 (88–124% compared to 172–243% in our main case, see Supplementary Fig. S1).
- We model a case where 90% of customers electrify by 2050 in Massachusetts instead of 80%, which results in higher overall increases in the cost of gas service from 2024–2050 (394–559%, compared to 172–243% in our main case, see Supplementary Fig. S1).
- We model a case where pipeline replacements are depreciated using units-of-production in Massachusetts, which results in an increase in the costs of gas service of 181–262% in 2050, compared to 172–243% in our main case (see Supplementary Fig. S3 and S4)
- We model a 2% annual escalation rate for the cost of pipeline replacements in Massachusetts, which results in an increase in the costs of gas service of 175–247% in 2050, compared to 172–243% without the escalation rate (see Supplementary Fig. S5)
- We model the energy burden impacts in Massachusetts after 2050, when more than 80% of households electrify (see Supplementary Fig. S6 and Supplementary Table S5 for further details on the energy burden distributions in Massachusetts).
- We model the increase in utility costs per customer in our national case using minimum, maximum, and average replacement costs per mile (see Supplementary Fig. S7).

- We model the energy burden impacts of income-based customer exit (from low to high income and vice versa) nationally (see Supplementary Fig. S8 and Supplementary Table S6 for further details on the energy burden distributions nationally).

# Availability of data and materials

The code and publicly available data to reproduce the analysis of this paper will be made available via an open-access repository (Zenodo) following publication.

# Acknowledgments

JGR was supported by the University of Wisconsin–Madison Office of Sustainability's Postdoctoral Research funding. MRE acknowledges support from a University of Wisconsin–Madison Vilas Associates Award.

We acknowledge Katherine Fisher and other members of HEET (Home Energy Efficiency Team) for their support with data curation about gas leaks and pipeline replacements in Massachusetts.

We also acknowledge Mark Kleinginna from HEET for helpful discussions to support our natural gas utility economic modeling.

# Author contributions

Conceptualization: M.R.E. and Z.M.; methodology: M.R.E., J.G.-R., and A.F.F.; formal analysis: J.G.-R., M.R.E., and A.F.F.; resources: M.R.E. and Z.M.; data curation: J.G.-R., and A.F.F.; writing – original draft: J.G.-R., M.R.E., and A.F.F.; writing – review and editing: all authors; visualization: J.G.-R.; supervision, M.R.E.; project administration: M.R.E.; funding acquisition: M.R.E.

# Ethics declarations

## Competing interests

The author(s) declare no competing interests.

# Additional information

See accompanying Supplementary Information.

# Supplementary information

## Supplementary figures

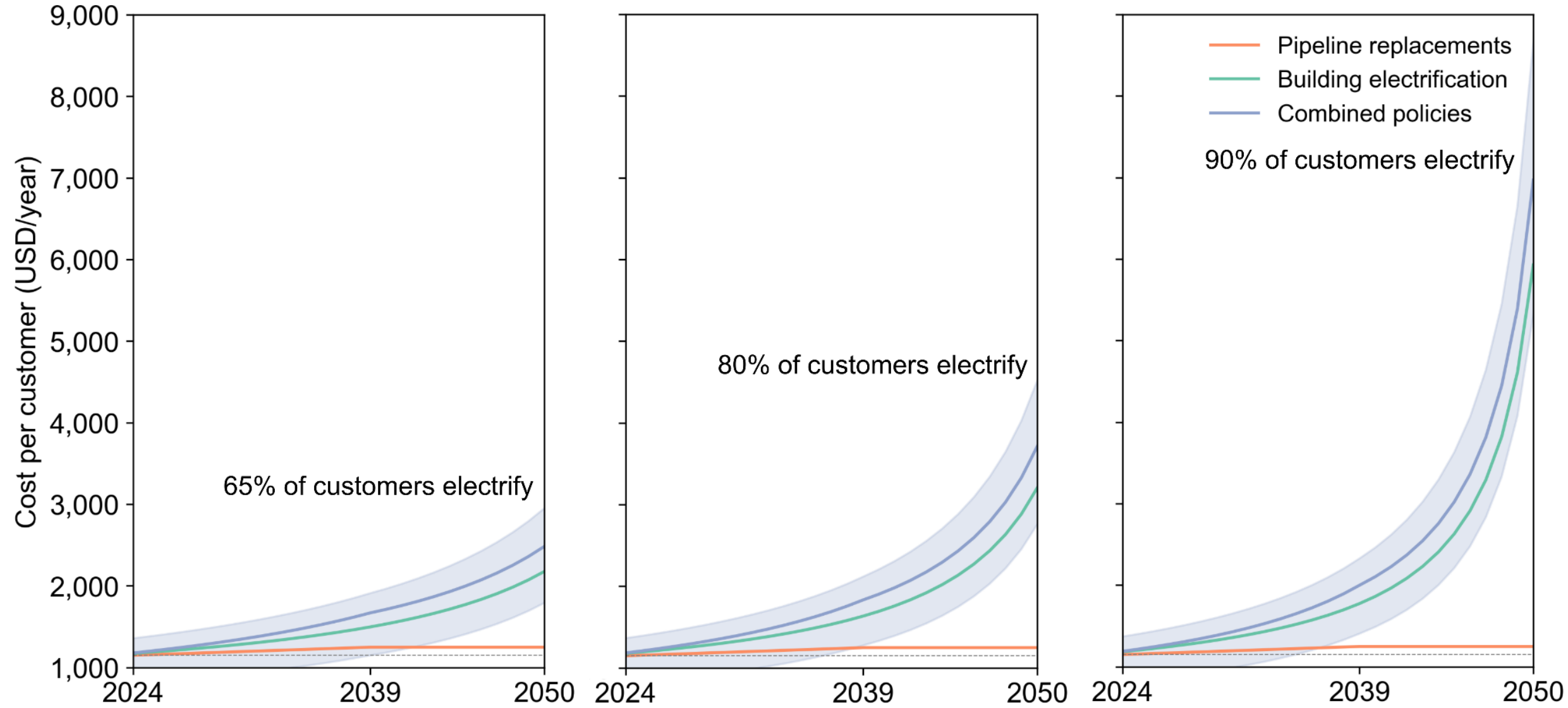

*Figure S1. Sensitivity analysis of average costs per customer for natural gas utilities under three scenarios: pipeline replacement, building electrification, and an uncoordinated transition that combines these two policies. Costs are compared to a baseline constant cost scenario (grey dotted line). The shaded area represents the range of costs across the six utilities in Massachusetts. Pipeline replacement scenarios use straight-line depreciation and an asset lifetime of 60 years. The left, middle, and right panes show the results when 65%, 80% (our main analysis), and 90% of customers electrify (compare to Fig. 3 in the main paper).*

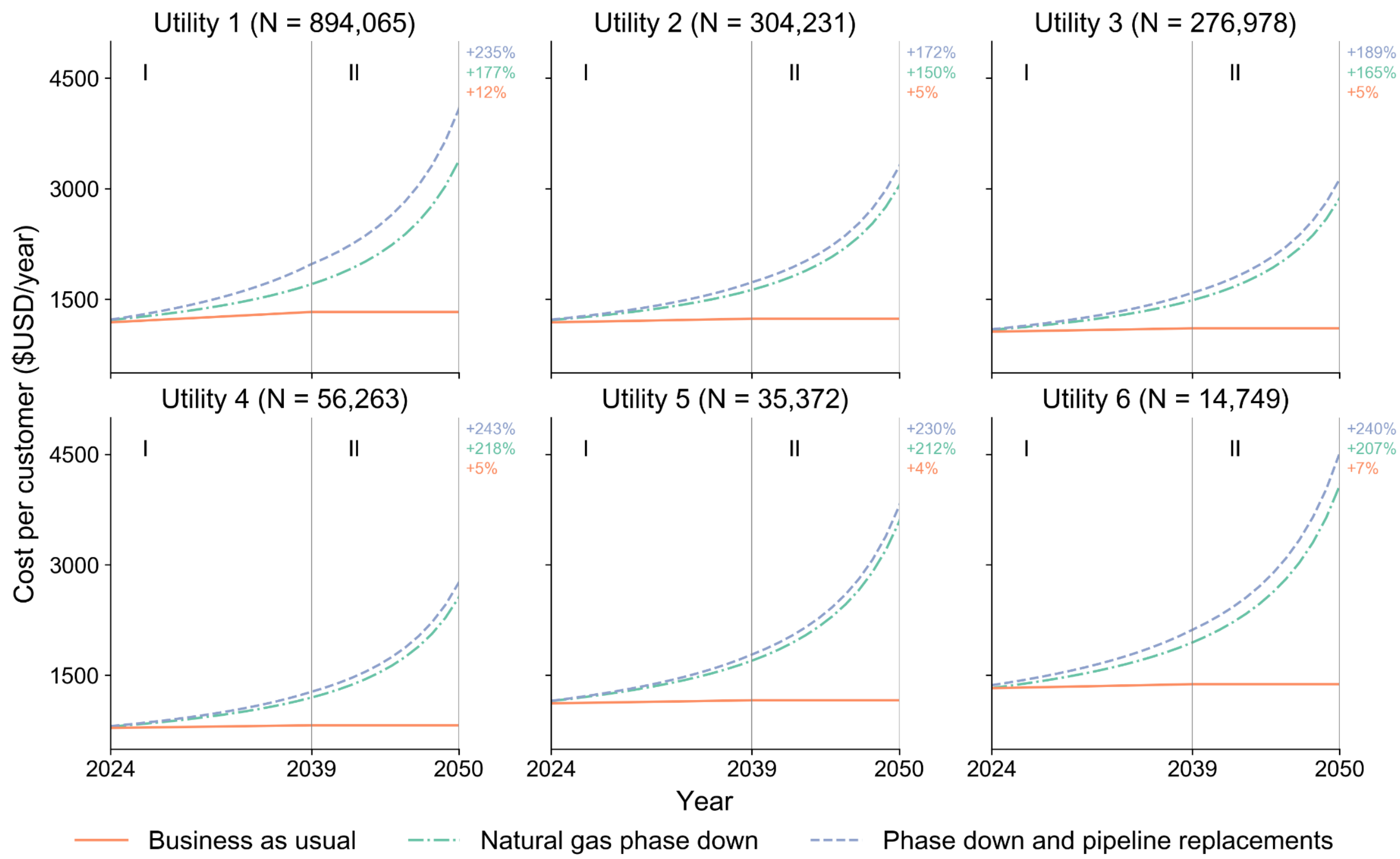

*Figure S2. Annual utility gas costs per customer with household heating transitions in all six major utilities in Massachusetts (see Fig. 3 in the main paper). The orange line shows the annual cost per customer pipeline replacement scenario. The green line shows the building electrification scenario, and the blue line shows an uncoordinated transition that combines these two policies.*

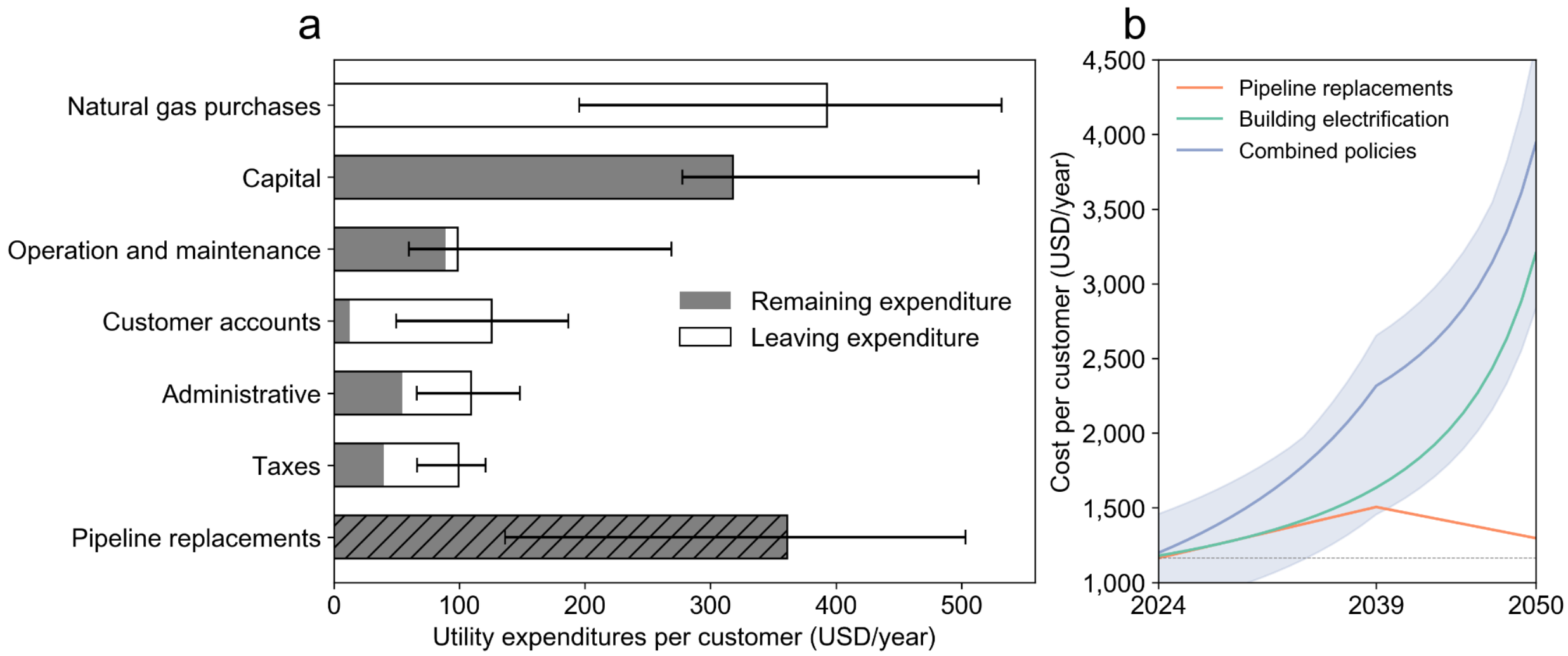

*Figure S3. Sensitivity analysis of the effects of an uncoordinated natural gas transition on natural gas costs in Massachusetts using units-of-production depreciation for pipeline replacements. (a) Customer weighted average costs per customer for each utility expenditure category. Pipeline replacement costs represent the annualized costs of all pipeline replacements by 2039. Error bars indicate the range of costs across all six major utilities in Massachusetts. (b) Average costs per customer for natural gas utilities under three scenarios: pipeline replacement, building electrification, and an uncoordinated transition that combines these two policies. Costs are compared to a baseline constant cost scenario (grey dotted line). The shaded area represents the range of costs across the six utilities in Massachusetts. Pipeline replacement scenarios use units-of-production depreciation, assuming gas is phased out by 2058.*

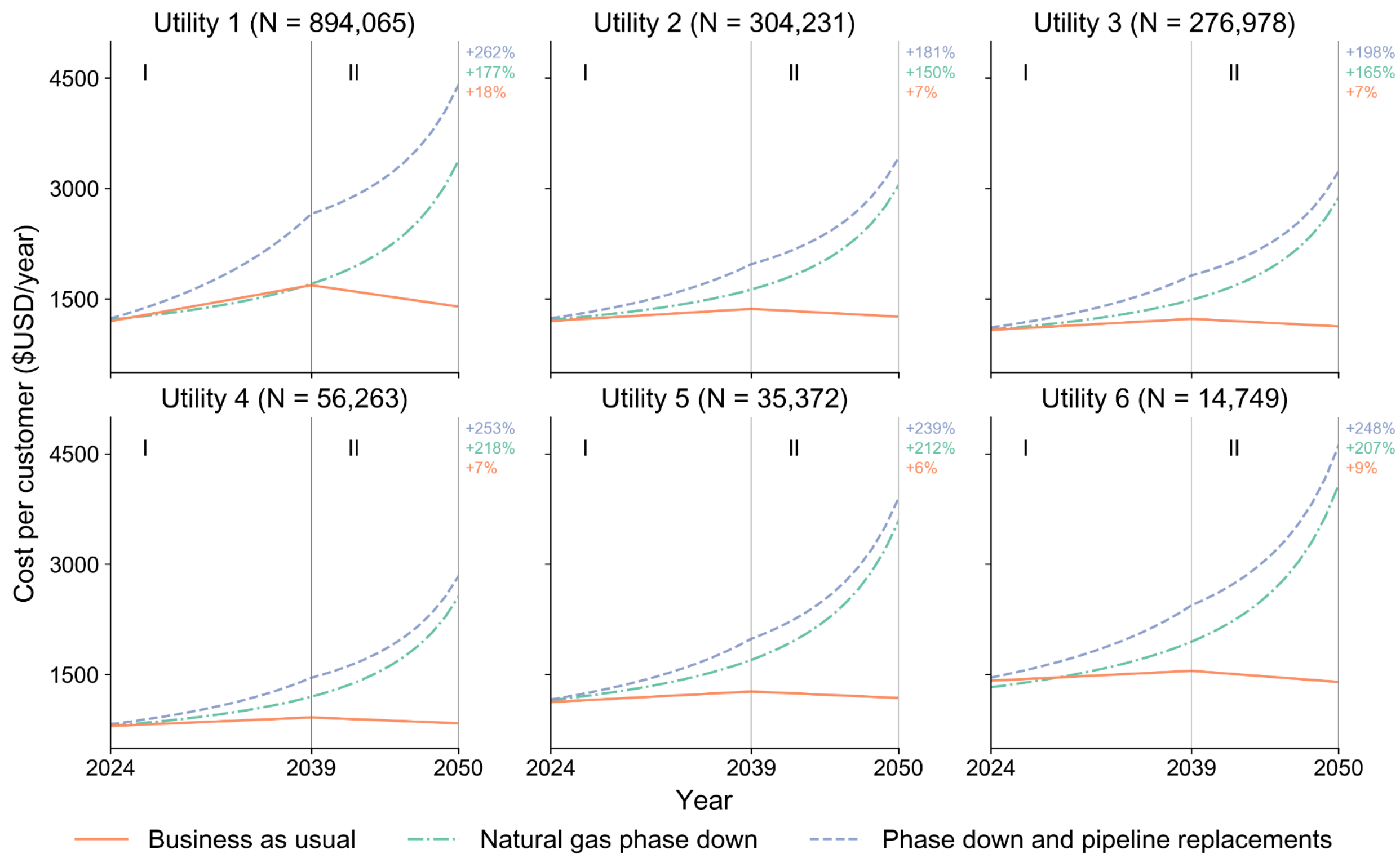

*Figure S4. Sensitivity analysis of annual utility gas costs per customer with household heating transitions in all six major utilities in Massachusetts using a units-of-production depreciation for pipeline replacements (see Fig. 3 in the main paper). The orange line shows the annual cost per customer pipeline replacement scenario. The green line shows the building electrification scenario, and the blue line shows an uncoordinated transition that combines these two policies.*

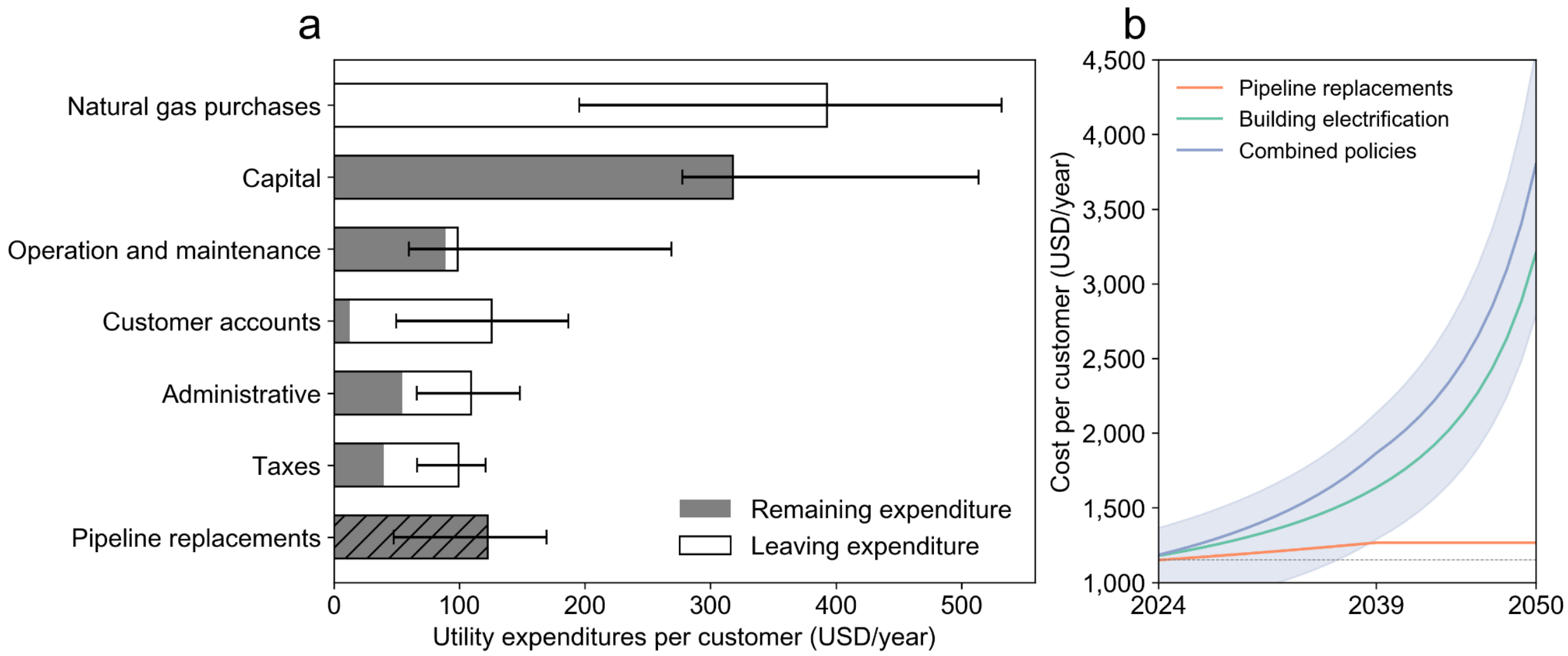

*Figure S5. Sensitivity analysis of the effects of an uncoordinated natural gas transition on natural gas costs in Massachusetts with a 2% cost escalation rate for pipeline replacements. (a) Customer weighted average costs per customer for each utility expenditure category. Pipeline replacement costs represent the annualized costs of all pipeline replacements by 2039. Error bars indicate the range of costs across all six major utilities in Massachusetts. (b) Average costs per customer for natural gas utilities under three scenarios: pipeline replacement, building electrification, and an uncoordinated transition that combines these two policies. Costs are compared to a baseline constant cost scenario (grey dotted line). The shaded area represents the range of costs across the six utilities in Massachusetts. Pipeline replacement scenarios use straight-line depreciation and an asset lifetime of 60 years.*

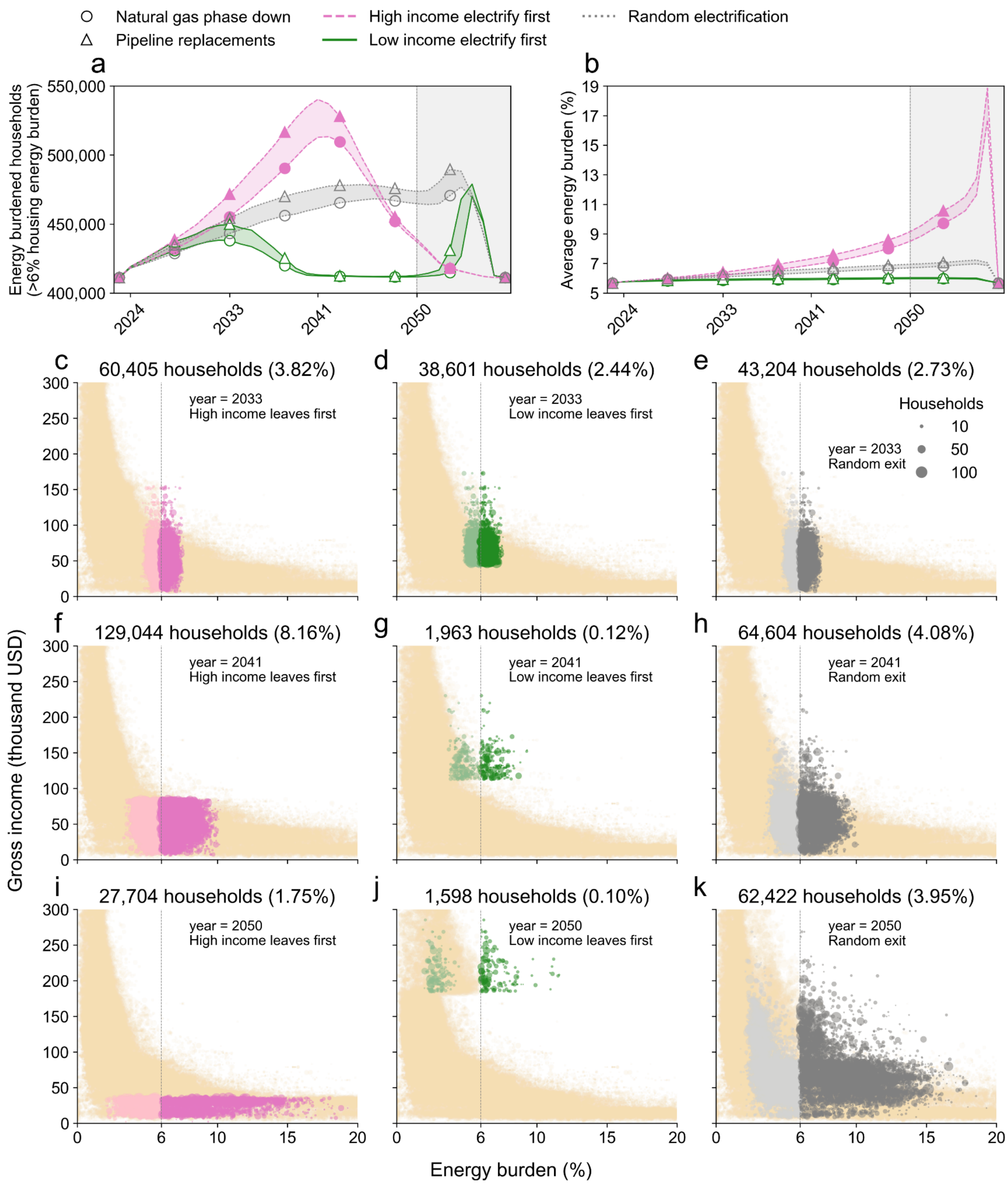

*Figure S6. Effects of an uncoordinated natural gas transition on energy burdens in Massachusetts through 2050 and after. (a) Number of energy burdened households over time under different electrification scenarios; the shaded area represents the additional impacts of pipeline replacement policies on burdens. (b) Corresponding average energy burdens. (c)–(k) Household-level energy burdens in 2033, 2041, and 2050. Dark colored bubbles represent households that become energy burdened; light colored bubbles show current energy burdens for these same households. Tan colored bubbles show households that do not cross the 6% energy burden threshold (i.e., their status as energy burdened is unchanged). Bubble size reflects the number of occupied housing units in the LEAD dataset.*

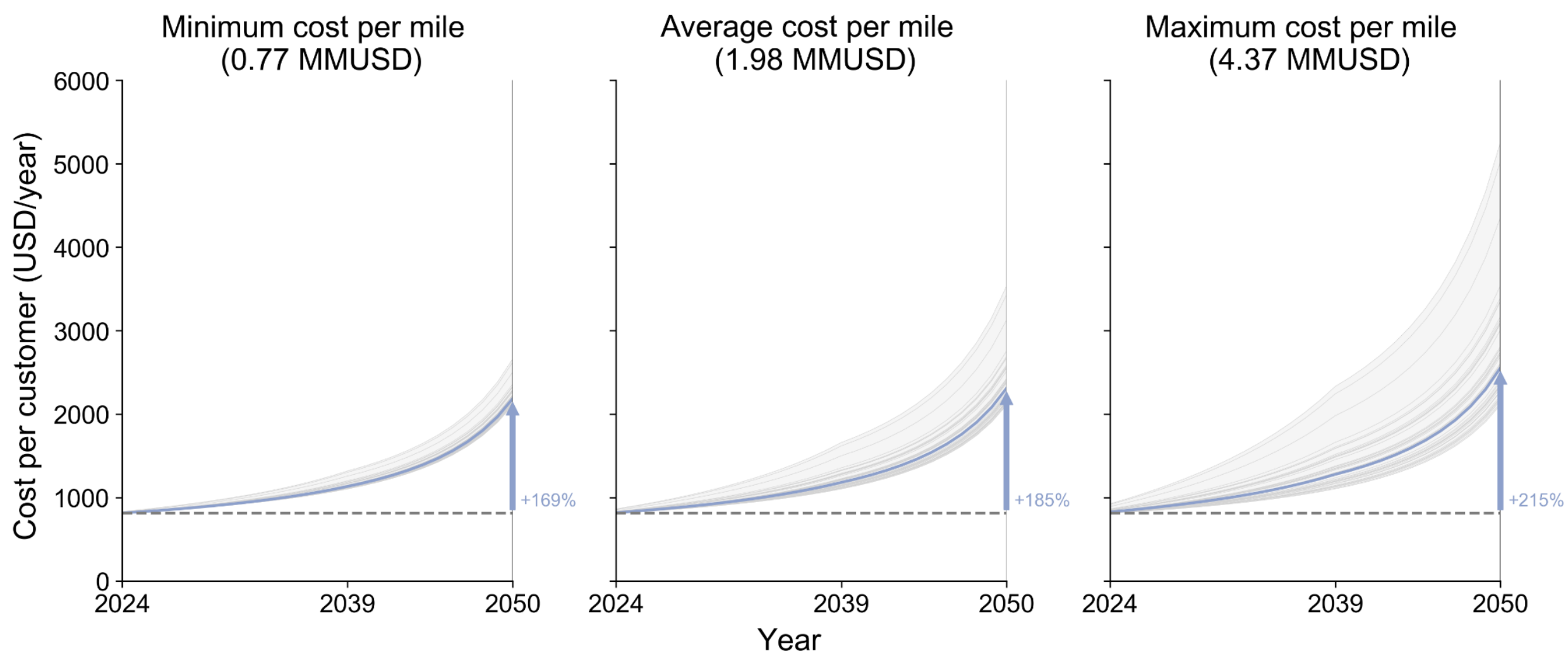

*Figure S7. Sensitivity analysis for utility costs per customer with natural gas phase down and pipeline replacement in 2050 using the minimum, average, and maximum cost-per-mile of pipeline replaced in our national data search (see Methods for details).*

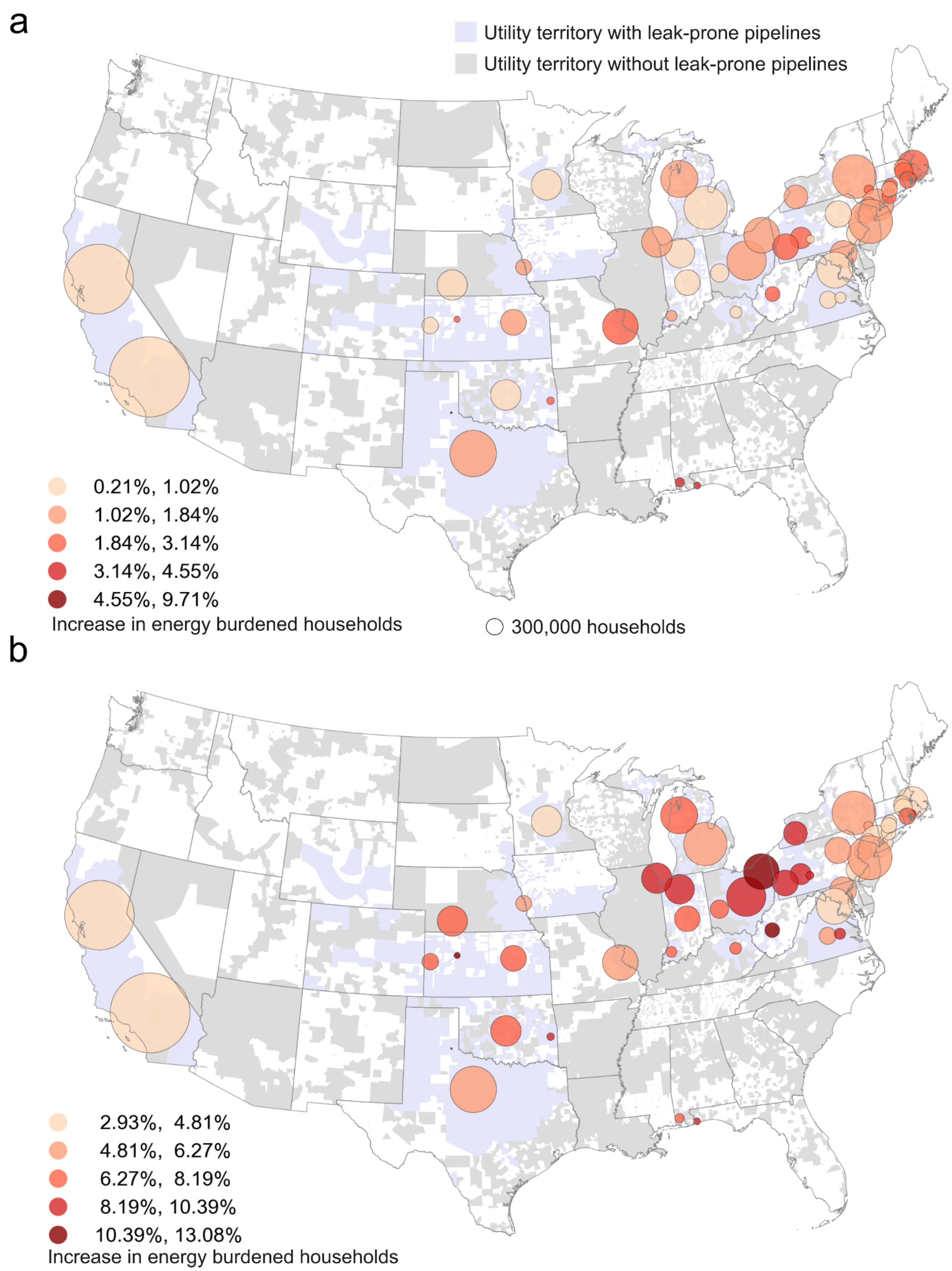

*Figure S8. Maps of the percentage increase of energy burdened households by 2050 when (a) low income customers electrify first and (b) vice versa (compare to Fig. 5d in the main paper). Gray zones in maps represent all gas utilities in the U.S., and violet zones represent the territories of the 50 utilities with leak-prone gas pipelines. Bubble size is proportional to the number of residential customers.*

# Supplementary tables

*Table S1. Average cumulative costs of electrification per customer in Massachusetts by 2050. The baseline scenario considers added costs of capital expenditures in pipeline replacements from 2024 to 2039 without electrification. (See Supplementary Fig. S4 for the units-of-production sensitivity analysis.)*

| Depreciation method for pipeline replacements | Baseline | Electrification without pipeline replacements | Electrification with pipeline replacements |
|---|---|---|---|
| Straight line depreciation (2024 USD) | 33,214 | 46,623 | 52,235 |
| Units of production (2024 USD) | 37,588 | 46,623 | 61,191 |

*Table S2. Pipeline replacement cost results from our national utility search. References are to utility websites or press releases (see Methods for details). Additional utility data we use to model our national case is shown in Table S3.*

| Utility | State(s) serviced | Pipeline replacement cost (2024 USD/mile) | Reference |
|---|---|---|---|
| 1 | NY | - | One-off projects (no evidence of replacement program)[1] |
| 2 | CA | - | One-off projects (no evidence of replacement program)[2] |
| 3 | OH | 1,083,742 | Confirmed active program[3] |
| 4 | NJ | 2,370,852 | Confirmed active program[4] |
| 5 | MA | - | Confirmed active program (missing cost data)[5] |
| 6 | TX | - | Confirmed active program (missing cost data)[6] |
| 7 | PA, WV | 1,480,485 | Confirmed active program[7] |
| 8 | OH | - | One-off projects (no evidence of replacement program)[8] |
| 9 | NY | - | Confirmed active program (missing cost data)[9] |
| 10 | MI | - | One-off projects (no evidence of replacement program[10] |
| 11 | WV | - | No evidence of active program[11] |
| 12 | NY | - | Potential inaccuracy in mileage and/or cost data[12] |
| 13 | IL | - | Potential inaccuracy in mileage and/or cost data[13] |
| 14 | MD | - | Confirmed active program (missing cost data)[14] |
| 15 | PA | - | One-off projects (no evidence of replacement program)[15] |
| 16 | RI | - | No evidence of active program[16] |
| 17 | PA | - | Confirmed active program (missing cost data)[17] |
| 18 | OK | - | Potential inaccuracy in mileage and/or cost data[18] |
| 19 | MD, PA, PA | 4,370,718 | Confirmed active program[19] |
| 20 | PA | 1,480,485 | Confirmed active program[20] |
| 21 | CT | - | No evidence of active program[21] |
| 22 | AL | - | No evidence of active program[22] |
| 23 | DC, MD, VA | - | Potential inaccuracy in mileage and/or cost data[23] |
| 24 | MA | - | Confirmed active program[24] |
| 25 | MA | - | Confirmed active program[25] |
| 26 | MI | 769,231 | Confirmed active program[26] |
| 27 | FL | - | With government funded project(s)[27] |
| 28 | AR, CO, IA, KS, NE, WY | - | One-off projects (no evidence of replacement program)[28] |
| 29 | KY | - | One-off projects (no evidence of replacement program)[29] |
| 30 | AR, OK | - | No evidence of active program [30] |
| 31 | CT | - | No evidence of active program[31] |
| 32 | MO | - | One-off projects (no evidence of replacement program)[32] |
| 33 | KS, OK | - | No evidence of active program [33] |
| 34 | CT | - | One-off projects (no evidence of replacement program)[34] |
| 35 | NJ, NJ, PA | - | No evidence of active program[35] |
| 36 | CO, KS | - | One-off projects (no evidence of replacement program)[36] |
| 37 | VA | - | With government funded project(s)[37] |
| 38 | NE | - | Confirmed active program (missing cost data)[38] |
| 39 | TX | - | No evidence of active program[39] |
| 40 | IN | - | Confirmed active program (missing cost data)[40] |
| 41 | VA, VA | - | No evidence of active program[41] |
| 42 | IN | 788,532 | Confirmed active program[42] |
| 43 | NY | - | No evidence of active program[43] |
| 44 | OH | - | No evidence of active program[44] |
| 45 | MA | - | Confirmed active program[45] |
| 46 | NY | - | No evidence of active program[46] |
| 47 | KS | - | With government funded project(s)[47] |
| 48 | IN | 788,532 | Confirmed active program[48] |
| 49 | MN | - | No evidence of active program[49] |
| 50 | CA | - | No evidence of active program[50] |

[1]https://www.nationalgridus.com/Long-Island-NY-Home/Natural-Gas-Safety/natural-gas-main-replacement
[2]https://www.socalgas.com/sustainability/pipeline-and-storage-safety/pipeline-replacement
[3]https://news.dominionenergy.com/2016-09-14-PUCO-Approves-Expansion-of-Dominion-East-Ohios-Pipeline-Infrastructure-Replacement-Program-Authorizes-Increase-in-Annual-Program-Investment-to-200-Million-by-2018
[4]https://nj.pseg.com/newsroom/newsrelease376
[5]https://www.nationalgridus.com/News/2023/11/National-Grid-Submits-Annual-Filing-for-Approval-of-Gas-Safety-Adjustment-Factors/#:~:text=Targeted%20electrification%20is%20a%20pillar,necessary%20to%20the%20electric%20infrastructure.
[6]https://s201.q4cdn.com/158157484/files/doc_presentation/2023/12/Aanlyst-Update-December-2023.pdf
[7]https://www.peoples-gas.com/news/archive/2018/07/replacing-pipelines.php
[8]https://www.columbiagasohio.com/services/work-in-your-neighborhood/replacement-projects

[9]https://investor.conedison.com/node/37026/pdf
[10]https://www.dteenergy.com/us/en/quicklinks/search/gas_map.html
[11]Unavailable website
[12]https://www.nationalfuel.com/corporate/safety/
[13]https://www.peoplesgasdelivery.com/services/main-replacement#:~:text=In%20total%2C%20our%20crews%20will,thawing%20cycles%20of%20Chicago's%20weather.
[14]https://www.bge.com/smart-energy/innovation-technology/reliability-improvements/natural-gas-investments
[15]https://www.columbiagaspa.com/services/work-in-your-neighborhood/pipeline-replacement-projects
[16]https://rhodeislandcurrent.com/2023/04/03/puc-approves-new-rhode-island-energy-capital-infrastructure-rehab-plan/
[17]https://www.puc.pa.gov/press-release/2015/puc-approves-modified-long-term-infrastructure-improvement-plan-for-peco-gas-allowing-faster-replacement-of-at-risk-mains
[18]https://s23.q4cdn.com/945657003/files/doc_financials/2024/sr/06/2024-sustainability-report-in-spreads-for-web-final.pdf
[19]https://ugi.reportablenews.com/pr/ugi-utilities-announces-2023-infrastructure-replacement-and-betterment-projects
[20]https://www.peoples-gas.com/news/archive/2018/07/replacing-pipelines.php
[21]Unavailable website
[22]Unavailable website
[23]https://www.washingtongas.com/-/media/f27bd5f1492b49da8234d123a9d6fd69.pdf
[24]https://www.mass.gov/info-details/gseps-pursuant-to-2014-gas-leaks-act
[25]https://www.mass.gov/info-details/gseps-pursuant-to-2014-gas-leaks-act
[26]https://www.consumersenergy.com/about-us/natural-gas-operations/pipeline-upgrade
[27]https://www.pensacolaenergy.com/community/news#:~:text=Pensacola%20Energy%20was%20awarded%20%2410,of%20Pensacola%20and%20Escambia%20County.
[28]https://www.blackhillsenergy.com/AuburnNE
[29]https://www.columbiagasky.com/services/work-in-your-neighborhood/pipeline-replacement-projects
[30]Unavailable website
[31]Unavailable website
[32]https://www.spireenergy.com/upgrades
[33]Unavailable website
[34]https://www.eversource.com/content/residential/safety/natural-gas-safety/system-management-maintenance/pipeline-integrity
[35]Unavailable website
[36]https://www.atmosenergy.com/pipeline-projects/colorado-service-territory/
[37]https://www.worldpipelines.com/regulations-and-standards/18062024/biden-harris-administration-announces-us200-million-in-grants-to-fix-ageing-natural-gas-pipes/
[38]https://www.mudomaha.com/news/infrastructure-replacement-program-2023-progress-report/
[39]Unavailable website
[40]https://www.in.gov/oucc/natural-gas/key-cases-by-utility/nipsco-gas-infrastructure-plans/#2020_2025_Plan
[41]Unavailable website
[42]https://wibc.com/360312/centerpoint-energy-continues-statewide-gas-pipeline-replacement-project/#:~:text=CenterPoint%20Energy's%20multi%2Dyear%20natural,wear%20and%20tear%20as%20quickly.
[43]Unavailable website
[44]Unavailable website
[45]https://massachusetts.libertyutilities.com/blackstone/libertys-infrastructure-updates-continue-throughout-spring-and-summer.html
[46]Unavailable website
[47]https://www.mwenergy.com/news/view/midwest-energy-awarded-8.6-million-in-grants-for-natural-gas-pipeline-replacement-equipment
[48]https://wibc.com/360312/centerpoint-energy-continues-statewide-gas-pipeline-replacement-project/#:~:text=CenterPoint%20Energy's%20multi%2Dyear%20natural,wear%20and%20tear%20as%20quickly.
[49]Unavailable website
[50]Unavailable website

*Table S3. Utility data for the Massachusetts case. Residential customer counts are taken from annual utility reports.[1] Pipeline replacement costs are derived from HEET data extraction of submitted replacement plans for 2024.[2] Total leak-prone pipeline miles are taken from Pipeline and Hazardous Materials Safety Administration (PHMSA), accessed November 1, 2024.[3]*

| Item | Utility 1 | Utility 2 | Utility 3 | Utility 4 | Utility 5 | Utility 6 |
|---|---|---|---|---|---|---|
| Residential customers | 894,065 | 304,231 | 276,978 | 56,263 | 35,372 | 14,749 |
| 2024 replacement cost (MMUSD) | 225.115 | 97.799 | 128.658 | 20.582 | 5.243 | 40.370 |
| 2024 pipeline replacements (miles) | 67.074 | 48.488 | 61.689 | 13.108 | 2.453 | 20.787 |
| 2024 cost per mile (MMUSD) | 3.356 | 2.017 | 2.086 | 1.570 | 2.138 | 1.942 |
| Leak-prone pipelines (miles) | 2093 | 443 | 364 | 76 | 42 | 30 |

*Table S4. Estimates of pipeline replacement costs for utilities across the U.S. Utility ID, state(s) serviced, and leak-prone miles are taken from PHMSA accessed November 1, 2024. Residential customer counts are extracted from the Energy Information Administration (EIA) form 176. Total replacement costs are calculated using the average cost-per-mile across our national data search and Massachusetts data (1.98 MMUSD, see Figure S4).*

| Utility | State(s) serviced | Residential customers | Leak-prone pipelines (miles) | Total replacement costs (MMUSD) | Replacement cost per customer (USD) |
|---|---|---|---|---|---|
| 1 | NY | 1,710,691 | 3,638 | 7,209 | 4,214 |
| 2 | CA | 5,711,670 | 3,369 | 6,675 | 1,169 |
| 3 | OH | 1,129,195 | 3,313 | 6,564 | 5,813 |
| 4 | NJ | 1,728,858 | 2,878 | 5,702 | 3,298 |
| 5 | MA | 877,015 | 2,201 | 4,361 | 4,972 |
| 6 | TX | 1,923,404 | 2,081 | 4,124 | 2,144 |
| 7 | PA, WV | 597,759 | 1,817 | 3,601 | 6,024 |
| 8 | OH | 1,376,619 | 1,564 | 3,098 | 2,250 |
| 9 | NY | 931,668 | 1,502 | 2,976 | 3,194 |
| 10 | MI | 1,225,639 | 1,452 | 2,876 | 2,347 |
| 11 | WV | 195,208 | 1,343 | 2,661 | 13,631 |
| 12 | NY | 505,863 | 1,180 | 2,339 | 4,624 |
| 13 | IL | 828,384 | 980 | 1,941 | 2,343 |
| 14 | MD | 655,373 | 945 | 1,872 | 2,856 |
| 15 | PA | 407,754 | 937 | 1,856 | 4,552 |
| 16 | RI | 247,508 | 731 | 1,448 | 5,851 |
| 17 | PA | 502,944 | 692 | 1,372 | 2,727 |
| 18 | OK | 834,915 | 677 | 1,342 | 1,607 |
| 19 | MD, PA, PA | 617,571 | 661 | 1,310 | 2,121 |
| 20 | PA | 58,603 | 643 | 1,274 | 21,732 |
| 21 | CT | 188,240 | 636 | 1,260 | 6,693 |
| 22 | AL | 79,483 | 590 | 1,168 | 14,700 |
| 23 | DC, MD, VA | 1,157,171 | 566 | 1,122 | 970 |
| 24 | MA | 301,577 | 482 | 956 | 3,169 |
| 25 | MA | 274,507 | 412 | 816 | 2,974 |
| 26 | MI | 1,679,546 | 351 | 696 | 415 |
| 27 | FL | 43,627 | 326 | 645 | 14,792 |
| 28 | AR, CO, IA, KS, NE, WY | 815,456 | 317 | 627 | 769 |
| 29 | KY | 123,979 | 292 | 579 | 4,670 |
| 30 | AR, OK | 50,728 | 269 | 533 | 10,504 |
| 31 | CT | 170,278 | 258 | 511 | 3,002 |
| 32 | MO | 1,125,735 | 252 | 499 | 443 |
| 33 | KS, OK | 591,987 | 241 | 477 | 805 |
| 34 | CT | 221,537 | 225 | 446 | 2,012 |
| 35 | NJ, NJ, PA | 283,501 | 186 | 368 | 1,296 |
| 36 | CO, KS | 244,825 | 133 | 263 | 1,074 |
| 37 | VA | 110,884 | 131 | 259 | 2,334 |
| 38 | NE | 223,456 | 128 | 253 | 1,132 |
| 39 | TX | 3,234 | 118 | 235 | 72,564 |
| 40 | IN | 786,344 | 106 | 210 | 267 |
| 41 | VA, VA | 263,114 | 104 | 206 | 784 |
| 42 | IN | 582,580 | 101 | 200 | 344 |
| 43 | NY | 128,682 | 90 | 179 | 1,390 |
| 44 | OH | 303,887 | 83 | 164 | 541 |
| 45 | MA | 56,197 | 82 | 162 | 2,888 |
| 46 | NY | 74,451 | 82 | 162 | 2,180 |
| 47 | KS | 32,986 | 80 | 159 | 4,809 |
| 48 | IN | 103,898 | 57 | 113 | 1,092 |
| 49 | MN | 829,979 | 53 | 104 | 126 |
| 50 | CA | 4,324,434 | 52 | 103 | 24 |

*Table S5. Energy burden impacts in Massachusetts with natural gas phasedown and pipeline replacement in 2033, 2041, and 2050. The total number of households is 1,581,658.*

| Case | High income leaves first (2041) | Low income leaves first (2033) | Random (2050) |
|---|---|---|---|
| Average energy burden (2024) | 5.68% | 5.68% | 5.68% |
| Average energy burden with phaseout | 6.89% | 5.85% | 6.70% |
| Average energy burden with replacements | 7.32% | 5.93% | 6.95% |
| Energy burdened households (2024) | 411,214 | 411,214 | 411,214 |
| Energy burdened households with phase down | 512,825 | 438,088 | 464,705 |
| Energy burdened households with replacements | 540,258 | 449,816 | 473,637 |
| % energy burdened households (2024) | 26.00% | 26.00% | 26.00% |
| % energy burdened households with phase down | 32.42% | 27.70% | 29.38% |
| % energy burdened households with replacements | 34.16% | 28.44% | 29.95% |

*Table S6. Energy burden impacts for the top 50 utilities with leak-prone infrastructure in the U.S. with natural gas phasedown and pipeline replacement. The total number of households is 37,238,689.*

| Case | High income leaves first (2044) | Low income leaves first (2030) | Random (2050) |
|---|---|---|---|
| Average energy burden (2024) | 4.74% | 4.74% | 4.74% |
| Average energy burden with phaseout | 5.73% | 4.84% | 5.37% |
| Average energy burden with replacements | 5.97% | 4.87% | 5.49% |
| Energy burdened households (2024) | 7,700,961 | 7,700,961 | 7,700,961 |
| Energy burdened households with phase down | 9,590,165 | 7,995,374 | 8,705,213 |
| Energy burdened households with replacements | 9,878,703 | 8,096,976 | 8,836,884 |
| % energy burdened households (2024) | 20.68% | 20.68% | 20.68% |
| % energy burdened households with phase down | 25.75% | 21.47% | 23.37% |
| % energy burdened households with replacements | 26.53% | 21.74% | 23.73% |

# Supplementary notes

**S1.** We calculate average energy burdens weighted by the number of occupied housing units in the LEAD Tool contains data for household energy expenditures (excluding transportation). We use the following equations to calculate our weighted average energy burdens:

$$\mathrm{EnergyBurden_h} = \frac{\mathrm{ElectricityExpenditures_h} + \mathrm{GasExpenditures_h} + \mathrm{OtherFuelExpenditures_h}}{\mathrm{HouseholdIncome_h}},$$

$$\mathrm{WeightedAverageEnergyBurden_h} = \sum_{\mathrm{h}=1}^{\mathrm{N_h}} \frac{\mathrm{w_h EnergyBurden_h}}{\mathrm{w_h}}.$$

In these equations, $N_h$ is the number of households, and $w_h$ is the number of occupied housing units in each household. The LEAD Tool uses a statistical representation of occupied housing units based on available census data from the American Community Survey (see Ma, O. et al. *Low-Income Energy Affordability Data (LEAD) Tool Methodology*. https://www.osti.gov/biblio/1545589 (2019) doi:10.2172/1545589).

**S2.** Note that the different sources provide different estimates of total pipeline replacements. For example, in our Massachusetts case, HEET data (collected from utility GSEP reports) estimates a total of 173 miles of pipelines were planned to be replaced in 2023, while PHMSA data indicates that 211 miles of leak-prone pipelines were replaced between 2022 and 2023.